\documentclass[reprint,prl,amsmath,amssymb,aps]{revtex4-1}

\usepackage{hyperref}
\usepackage{comment}
\hypersetup{colorlinks,urlcolor=blue}
\usepackage{graphicx}
\usepackage[percent]{overpic}
\usepackage{dcolumn}
\usepackage{bm}

\begin{document}
\preprint{APS/123-QED}
\title{Unsupervised Machine Learning of Quenched Gauge Symmetries: \\ A Proof-of-Concept Demonstration}

\author{Daniel Lozano-G\'omez}
\altaffiliation[These authors contributed equal work.]{}
 
\author{Darren Pereira}
\altaffiliation[These authors contributed equal work.]{}
 
\author{Michel J.P. Gingras}
 
\affiliation{Department of Physics and Astronomy, University of Waterloo, Ontario, N2L 3G1, Canada }

\begin{abstract}
In condensed matter physics, one of the goals of machine learning is the classification of phases of matter. The consideration of a system's symmetries can significantly assist the machine in this goal. We demonstrate the ability of an unsupervised machine learning protocol, the Principal Component Analysis method, to detect hidden quenched gauge symmetries introduced via the so-called Mattis gauge transformation. Our work reveals that unsupervised machine learning can identify hidden properties of a model and may therefore provide new insights into the models themselves.
\end{abstract}

\maketitle

\textit{Introduction} --  Machine learning (ML) has in recent years proven to be a powerful pattern recognition tool with applications in various branches of science. These techniques have shown their ability to extract, identify, and even propose descriptive patterns found in the input data. Particularly in condensed matter physics, the application of ML techniques began with the use of the Principal Component Analysis (PCA) method \cite{WangUnsupervisedIsing} and neural networks \cite{MelkoIsingNN} to identify the ferromagnetic and paramagnetic phases of the Ising model on a square lattice. Since then, this field has exploded with a variety of ML applications \cite{MehtaRev,DunjkoRev,MLinPhysicsReview2019}. These techniques and applications can be broadly grouped into two categories: supervised ML (SML), in which the input data is labelled to train the machine \cite{KaoReinforceMC,GreitemannSVM,BeachMLVortices,GreitemannHiddenOrder, BeachMLVortices,Ponte_kernel,Pollet_svm, PhysRevB.100.224202, PhysRevLett.122.200401,PhysRevB.98.104426, PhysRevA.95.012335, PhysRevE.100.050102,2019npjQI,2019NatPh15,PhysRevE.99.023304,PhysRevB.100.045129,WetzelSU2}; and unsupervised ML (UML), in which the input data is unlabelled and the machine proposes its own classification scheme \cite{WangZhaiPCA1,WangZhaiPCA2,SinghUMLXY,WetzelAutoencoders, Zhang_percolation,2019npjCM5,PhysRevB.96.205146,PhysRevE.99.023304,PhysRevB.100.045129,WetzelSU2, wu2019generalized,hou2019minimal,2018JChPh.149s4109J,2019JPhCo3g5006S}. As a major task of the condensed matter physicist, the classification of phases in various models has remained central among these applications. Evidence is accumulating that the machine learning of phases can be guided by physical insights into the model or system, such as symmetries. This has been most clearly demonstrated by exploiting properties such as locality and translational symmetry via convolutional neural networks \cite{MelkoIsingNN}, or by taking advantage of symmetry-breaking to extract order parameters for hidden orders \cite{GreitemannSVM}. 

In light of the benefits that these physically-inspired shortcuts provide, one may ask a question of foremost importance for the usage of ML in physics: is it possible for ML to provide theoretical insight into the hidden or unknown properties \emph{of a model itself}? A fitting testing ground for such a question is physical models possessing gauge symmetries, as these models can be simplified by a suitable mathematical transformation. Our question then becomes a matter of determining if ML can detect the gauge symmetry of these models without prior knowledge. Doing so would prove that ML is capable of learning fundamental mathematical details of the studied model and not just thermodynamic quantities. This ability offers clear benefits for various branches of physics, including the aforementioned exploitation of symmetries for phase classification. Furthermore, the controlled mathematical nature of these gauge-symmetric models would also suggest their use as a probe of how ML methods work and what they are truly learning.

To explore this question, we therefore require (i) a model that seems complex but can be simplified by some gauge transformation, and (ii) a UML method whose self-determined classification scheme can be exposed. In light of (i), we study the Mattis Ising Spin Glass (MISG) \cite{Mattis,FischerSpinGlass} and the Mattis XY Gauge Glass (MXYGG) models \cite{FischerSpinGlass}. At first glance, the MISG and MXYGG models look prohibitively complex: the Hamiltonians for these models possess almost arbitrary bond interactions, which make an analytical approach seem intractable. A visual snapshot of their ground state configurations displays no recognizable pattern but instead appears completely disordered. However, the MISG and MXYGG models can be transformed into the regular ferromagnetic Ising and $XY$ models, respectively, under a (Mattis) gauge transformation \cite{FischerSpinGlass}. Regarding (ii), an important consideration is the trade-off between interpretability and scalability. We therefore use PCA \cite{KarlPeason}, which is highly interpretable and simple to apply, as opposed to neural-network-based methods, which may be more powerful but are not as open to interpretation.

The outline of the paper is as follows. We first describe the MISG and MXYGG models, as well as the Mattis gauge transformation, in the Models section. We then give a brief introduction to PCA in the Methods section. In the Results section, we demonstrate that PCA \textit{is} able to identify the gauge variables that quantify the Mattis gauge transformation. PCA additionally finds that the bond-disordered MISG and MXYGG models are simply disguised versions of the regular Ising and $XY$ models, classifying the phases in the former gauge-transformed models in exactly the same manner as it would with the regular models. Our work suggests that interpretable ML methods can therefore be used to reveal hidden features in the models themselves, giving a positive answer to our above question. We conclude by discussing the implications of our findings for investigations into other models and for other ML applications.
\newline

\textit{Models} -- The MISG model \cite{Mattis,FischerSpinGlass} on a square lattice is defined by the Hamiltonian 
\begin{equation}
 H=-\sum_{\langle i,j\rangle}J_{ij}\sigma_i^z\sigma_j^z,
\label{eq:Ising-M}
\end{equation}
where the spin variables are $\sigma_i^z=\pm 1$. The couplings $\{J_{ij}\}$ are free to take the values $\pm J$ randomly, with the \emph{imposed constraint} that the product $P$ of the couplings around a square plaquette is positive:
\begin{equation}
  P\equiv\prod_{\langle i,j\rangle \in \square} J_{ij}\textrm{ \textgreater 0}.\label{eq:plaquette_constraint}
 \end{equation}
This constraint enforces a non-frustrated ground state in the system and allows a so-called Mattis gauge transformation to be applied \cite{Mattis,FischerSpinGlass}. This gauge transformation reexpresses the interaction couplings as $J_{ij}=\epsilon_i\epsilon_j J$, where $\{\epsilon_i\}$ are random site (gauge) variables that take values of $\pm 1$. Through this transformation, the Hamiltonian \eqref{eq:Ising-M} becomes
\begin{equation}
 H=-J\sum_{\langle i,j\rangle}\epsilon_i\sigma_i^z \epsilon_j\sigma_j^z=-J\sum_{\langle i,j\rangle}\tau_i^z \tau_j^z,
\end{equation}
where  $\tau_i^z\equiv\epsilon_i \sigma_i^z= \pm1$ are new Ising variables. 

It is now clear that this system possesses a well-defined order parameter given by the Ising model ``$\tau-$magnetization'', 
\begin{equation}
     M\equiv \langle\sum_i\tau_i^z\rangle=\langle\sum_i\epsilon_i \sigma_i^z\rangle,
     \label{eq:magnetization_ising}
 \end{equation}
illustrating that the MISG model is nothing but an Ising model in disguise. Further information about this mapping is given in the Appendix. 

Similarly, the MXYGG model is described by an $XY$ model with random phase factors $\{A_{ij}\}$ \cite{Michel1,Michel2,Michel3},
\begin{equation}
H=-J\sum_{\langle i,j\rangle} \cos(\Delta \phi_{ij}-A_{ij}),\label{eq:xy-M}
\end{equation}
 where $\Delta \phi_{ij}=\phi_{i}-\phi_{j}$ is the difference between the on-site angular variables $\phi_i \in [0, 2\pi)$. This is equivalent to an $XY$ model with random Heisenberg exchange $J_{ij} \equiv J\cos(A_{ij})$ and Dzyaloshinsky-Moriya interactions $D_{ij} \equiv J\sin(A_{ij})$.

 This Hamiltonian is unfrustrated as long as the phase factors around a plaquette add to a multiple of $2\pi$,  \textit{i.e.} $ P_{XY}=(\sum_{\langle i,j\rangle\in \square} A_{ij})\textrm{ mod }2\pi= 0$ \cite{FischerSpinGlass}. A Mattis gauge transformation can then be applied by defining random site (gauge) variables $\{b_i\}$ such that $A_{ij}=b_i-b_j$, with $b_i \in [0, 2\pi)$. The Hamiltonian then becomes
\begin{equation}
H=-J\sum_{\langle i,j\rangle} \cos(\Delta \theta_{ij}),
\end{equation}
where $\theta_i \equiv \phi_i + b_i$ are new $XY$ variables. The MXYGG model can thus be mapped onto a ferromagnetic $XY$ model under this gauge transformation. This model possesses a  magnetization vector $\bm{M} = \langle \sum_i (\cos \theta_i,\sin \theta_i) \rangle$, or
\begin{eqnarray}
\begin{split}
M_x&= \langle \sum_i \left(\cos{\phi_i}\cos{b_i} - \sin{\phi_i}\sin{b_i}\right) \rangle, \\
M_y&= \langle \sum_i \left(\sin{\phi_i}\cos{b_i} + \cos{\phi_i}\sin{b_i}\right) \rangle.\\ 
\end{split}\label{eq:XYMag}
\end{eqnarray}

\textit{Methods} -- PCA is a dimensional reduction technique that identifies which linear combinations of the input data best characterize the full dataset. The input data for this method is defined as $n$ sets of configurations $\{x_i(T_j)\}$ of an $N-$site system, where $x_i$ is some variable (e.g. $\sigma_i^z$) associated with the $i^\textrm{th}$ site and sampled at a temperature $T_j$ ($j = 1,\ \dots,\ n$). The full dataset can then be formatted as a data matrix $X_{\mathrm{data}}$,
\begin{equation}
    X_{\mathrm{data}}\equiv\begin{pmatrix}
    \{x_i(T_1)\}\\
    \{x_i(T_2)\}\\
    \vdots\\
     \{x_i(T_n)\}\\
    \end{pmatrix}.
\end{equation}
After each row is centered by subtracting its mean value, the covariance matrix defined as $X_{\textrm{data}}^{T}X_{\textrm{data}}$ is diagonalized. The normalized eigenvalues and eigenvectors obtained are the so-called explained variance ratios $\{\lambda_k\}$ and principal components $\{\vec{u}^{(k)}\}$, respectively. Note that the eigenvectors can be rescaled by any convenient factor, such as the system size. The projection $\ell^{(k)}(T_j)$ of the $j^\textrm{th}$ configuration $\{x_i(T_j)\}$ onto the $k^\textrm{th}$ principal component $\vec{u}^{(k)}$ takes the form
\begin{equation}
    \ell^{(k)}(T_j) \equiv \sum_i u^{(k)}_{i} x_i(T_j).
    \label{eq:projection}
\end{equation}
These linear combinations $\{\ell^{(k)}(T_j)\}$ are the new quantities used by PCA to characterize the full dataset, where their relative importance is given by the values of their explained variance ratios. By construction, the $i^\textrm{th}$ value of any principal component is the coefficient multiplying the variable $x_i$ for any projection $\ell^{(k)}(T_j)$; therefore, the components of the eigenvector $\vec{u}^{(k)}$ directly contain site-dependent information. Plotting different projections against each other visually reveals how PCA ``clusters'' the input data and along which projections the data is most or least correlated. These clusters are composed of points which represent configurations with similar values of the projections.

PCA is applied to the MISG model by using spin configurations $\{\sigma_i^z(T_j)\}$ as the input data $\{x_i(T_j)\}$, with $T_j\in [J, 4J]$ sampled from a single spin flip Monte Carlo (MC) algorithm of a system of $N = 2500$ spins ($L=50$). The $\{\epsilon_i\}$ gauge variables at each site are randomly chosen as either $\pm 1$ with equal probabilities. For the MXYGG model, a MC simulation is performed in a system of $900$ spins ($L=30$) for temperatures $T_j\in [0.2J,1.8J]$ to sample the continuous angular variables $\{\phi_i\}$. PCA is then applied to three different datasets: $\{\cos(\phi_i)(T_j)\}$ (the ``$X$ dataset''), $\{\sin(\phi_i)(T_j)\}$ (the ``$Y$ dataset''), or $\{\{\cos(\phi_i)(T_j)\},\{\sin(\phi_i)(T_j)\}\}$ (the full dataset). The gauge variables $\{b_i\}$ are randomly drawn from a discrete distribution $\{\frac{2\pi n}{5}\mid n=1, \dots,5\}$ \footnote{It is easier to determine PCA's success with determining the gauge variables if they are taken from a discrete distribution, which can be clearly identified in a histogram such as Fig.~\ref{fig:extracted}, as opposed to a continuous distribution.}. In order to sample uncorrelated data, $3\times 10^4$ thermalization sweeps and $5\times10^4$ measurement sweeps are used at every temperature for both models; 50 different temperatures are selected. Sampling is done every $50$ (100) measurement sweeps for the MISG (MXYGG) model, producing $n = 5\times 10^4$ ($n = 2.5\times 10^4$) configurations. In both the MISG and MXYGG cases, PCA has no information about the gauge variables $\{\epsilon_i\}$ or $\{b_i\}$, nor the Ising variables $\{\tau_i^z\}$ or $XY$ variables $\{\theta_i\}$. The objective is to determine if PCA can identify these gauge variables and the underlying Ising or $XY$ models regardless, to which we now turn. 
\newline

\textit{Results for the MISG Model} -- PCA is applied to the configurations sampled through MC simulations for the MISG model. Plotting $\ell^{(1)}$ versus $\ell^{(2)}$ reveals a central high-temperature cluster and two adjacent low-temperature clusters as illustrated in Fig.~\ref{fig:PCA_Ising} of the Appendix, which is precisely how PCA clusters the input data of the regular ferromagnetic Ising model \cite{WangUnsupervisedIsing}. The similarity in this clustering suggests that PCA is characterizing the input data according to the Ising magnetization order parameter as it did in the regular case, thereby detecting the underlying Ising model. 

\begin{figure}
    \begin{overpic}[width=\columnwidth]{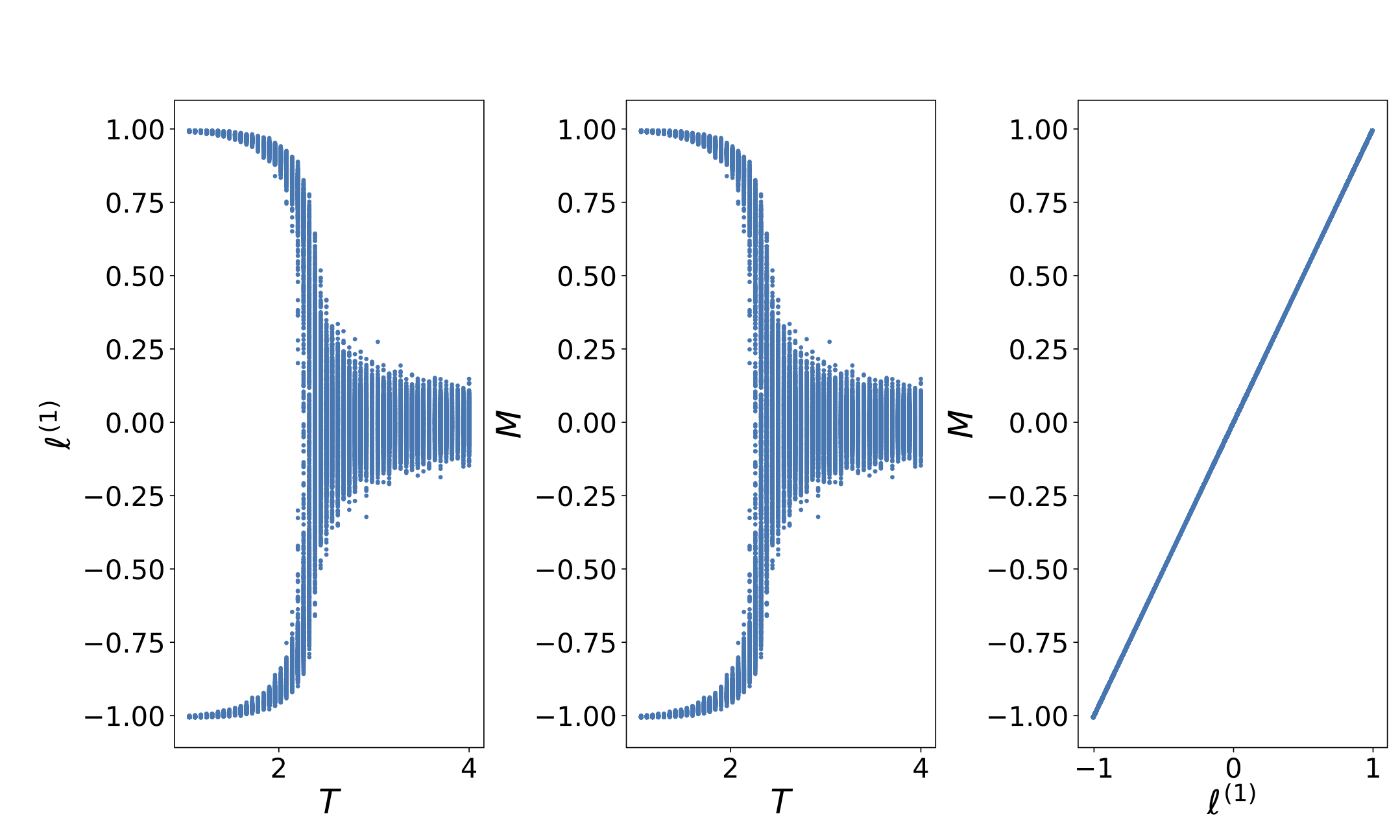}
    \put(20.0,55){(a)}
    \put(52,55){(b)}
    \put(84.5,55){(c)}
    \end{overpic}
    \caption{Comparison of (a) $\ell^{(1)}(T)$, the projection of the spin configuration data $\{\sigma_i^z\}$ onto the first principal component, with (b) the $\tau-$magnetization $M$ of the MISG model from MC simulations. (c) Plot of $\ell^{(1)}(T)$ against $M$.
    }
    \label{fig:Mattis_PCA}
\end{figure}

\begin{figure}
    \begin{overpic}[width=\columnwidth]{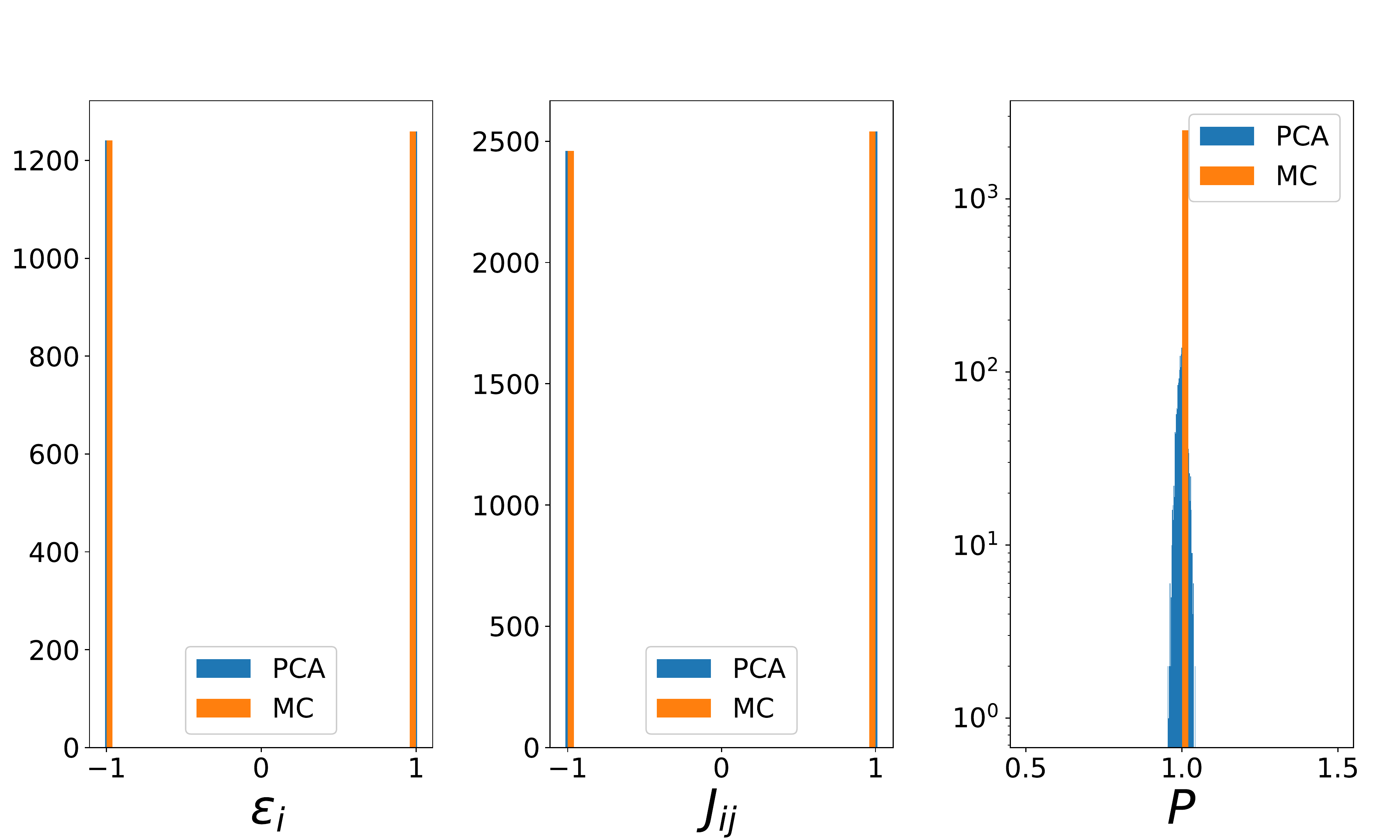}
    \put(16.5,55){(a)}
    \put(49,55){(b)}
    \put(82.5,55){(c)}
    \end{overpic}
    \caption{Histograms of the (a) gauge variables $\{\epsilon_i\}$, (b) bond interactions $\{J_{ij}\}$, and (c) square plaquette values $\{P\}$ for the MISG model, comparing real values from MC (known) and values from PCA (learned).}
    \label{fig:plaquettes_pca_mattis_reg}
\end{figure}

To verify this quantitatively, the projection $\ell^{(1)}$ of the input data onto the first (and most important) principal component is compared with the $\tau$-magnetization calculated within MC simulations using Eq.~\eqref{eq:magnetization_ising}, as shown in Fig.~\ref{fig:Mattis_PCA}. The resemblance of the projection in Fig.~\ref{fig:Mattis_PCA}a to the Ising magnetization in Fig.~\ref{fig:Mattis_PCA}b demonstrates that PCA is learning this order parameter. This is further confirmed when this projection is plotted against the Ising magnetization in Fig.~\ref{fig:Mattis_PCA}c, revealing a linear relationship with a slope of 1.
Since $\ell^{(1)}$ is equivalent to the $\tau$-magnetization even when PCA was only provided with $\{\sigma_i^z\}$, the first principal component contains information about the gauge variables $\{\epsilon_i\}$ that are hidden from PCA. In other words, the components of $\vec{u}^{(1)}$ are identified as the values of the gauge variables in the lattice. Therefore, by comparing the projection $\ell^{(1)}$ with Eq. \eqref{eq:magnetization_ising}, the \emph{learned} gauge variables can be extracted. Moreover, since PCA provides site-dependent information, the learned bond interactions $\{J_{ij}\}$ and square plaquette values $\{P\}$ can be calculated and compared with the original values. Histograms of the learned and known gauge variables $\{\epsilon_i\}$, bond interactions $\{J_{ij}\}$ and square plaquette values $\{P\}$ values are illustrated in Fig~\ref{fig:plaquettes_pca_mattis_reg}.
As can be seen in Fig.~\ref{fig:plaquettes_pca_mattis_reg}a and Fig.~\ref{fig:plaquettes_pca_mattis_reg}b, the learned values of the gauge variables and the bond interactions are described by a bimodal distribution centered around $\pm1$. These distributions agree with the ones produced with the known gauge variables used in the MC simulation. Moreover, a remarkable result comes from the distribution for the square plaquette values $\{P\}$, shown in Fig.~\ref{fig:plaquettes_pca_mattis_reg}c: this distribution is centered near the value $P=1$ which defines the plaquette constraint used in the MC simulation. PCA's ability to learn the values of the gauge variables is additionally provided by MC simulations: when the learned gauge variables $\{ \epsilon_i \}$ are used within MC simulations, the resulting energy per spin and specific heat curves are equivalent to the original curves which used the known gauge variables, as detailed in Fig.~\ref{fig:MCRerun} of the Appendix. 

\begin{figure}
    \begin{overpic}[width=\columnwidth]{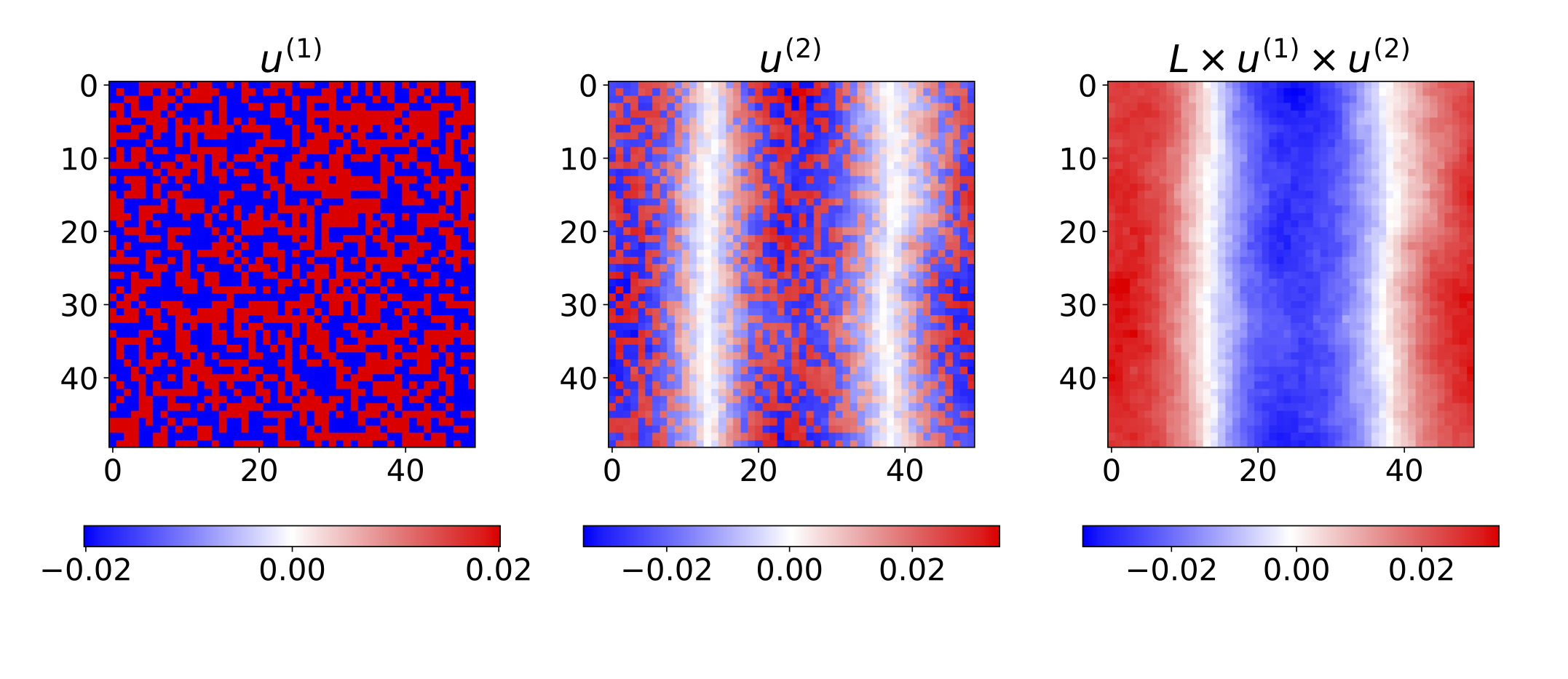}
    \put(17,45){(a)}
    \put(48,45){(b)}
    \put(80,45){(c)}
    \end{overpic}
    \caption{Values of the (a) first and (b) second principal components of the MISG model, plotted on the lattice sites that they are associated with. (c) Product of the values of the first and second principal components on each site. This plot has been rescaled by the lattice dimension $L$.}
    \label{fig:second_pc_gauged}
\end{figure}

After extracting the gauge variables, we can apply this learned gauge transformation to other quantities to confirm that the MISG model is transformed into the regular Ising model. For example, the second principal component of this model has been computed and plotted on the associated lattice sites \cite{SinghUMLXY}; by multiplying the second principal component of the MISG model by the first, as shown in Fig.~\ref{fig:second_pc_gauged}, the known regular Ising result is reconstructed. This operation is therefore equivalent to applying the gauge transformation to go from the MISG model to the regular Ising model. Altogether, this comparison of the learned and known gauge variables and thermodynamic quantities demonstrates PCA's ability to identify the correct values of these quenched gauge variables. 

\begin{figure}
    \centering
    \begin{overpic}[width=\columnwidth]{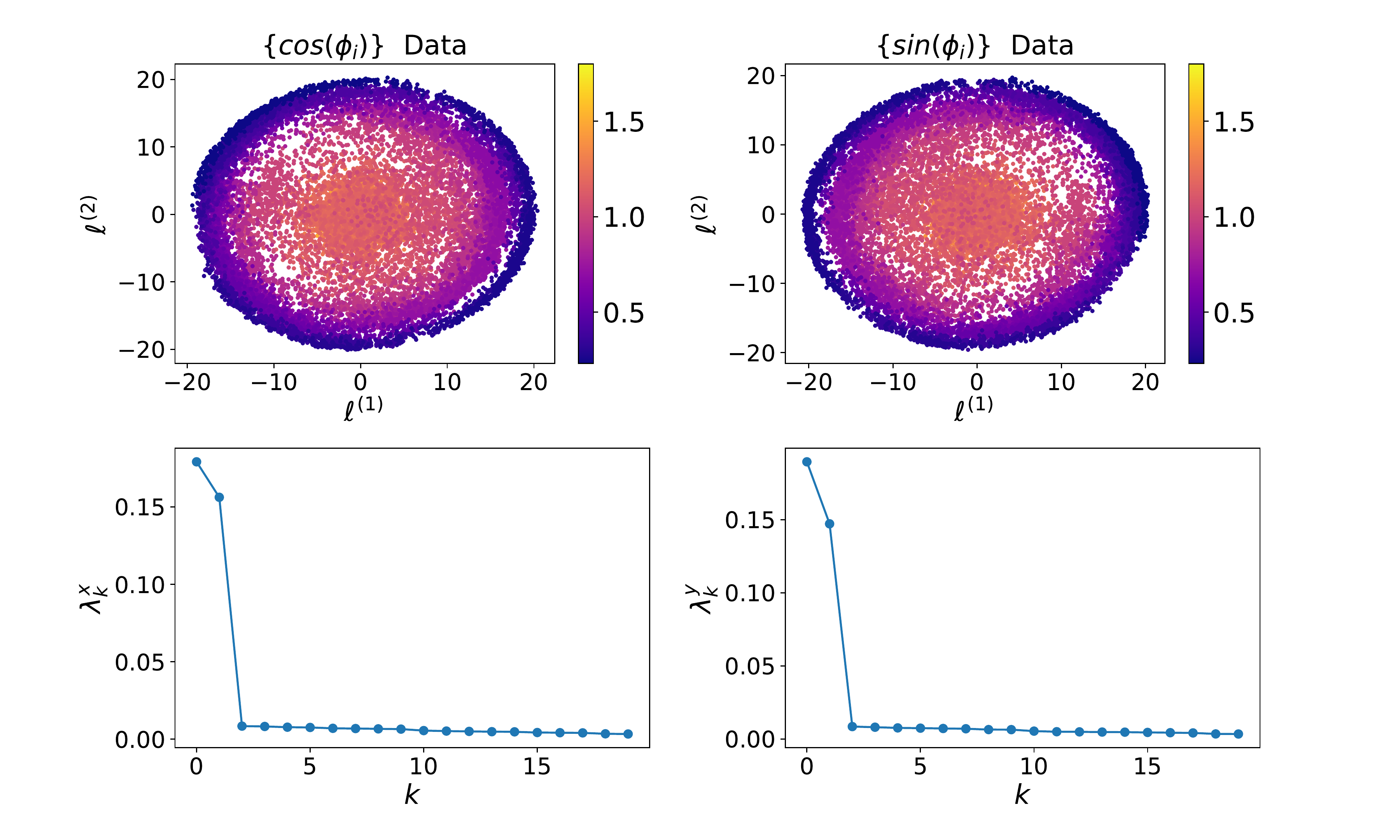}
    \put(35,53){(a)}
    \put(78,53){(b)}
    \put(40,24){(c)}
    \put(84,24){(d)}
    \end{overpic}
    \caption{Principal component projections $\ell^{(1)}$ versus $\ell^{(2)}$ ((a) and (b)) and explained variance ratios ((c) and (d)) for the MXYGG model, for PCA applied to $\{\cos{(\phi_i)}\}$ or $\{\sin{(\phi_i)}\}$ only.}
    \label{fig:gauge}
\end{figure}
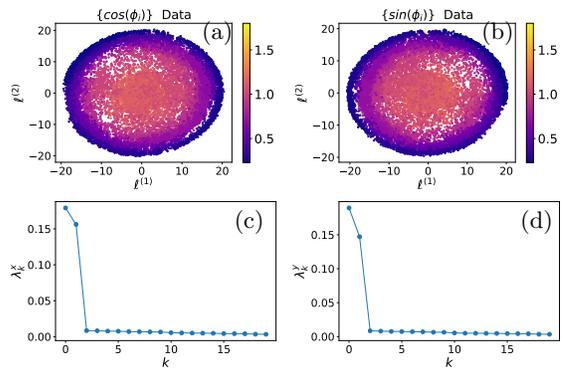

\begin{figure}
\centering
    \includegraphics[width=\columnwidth]{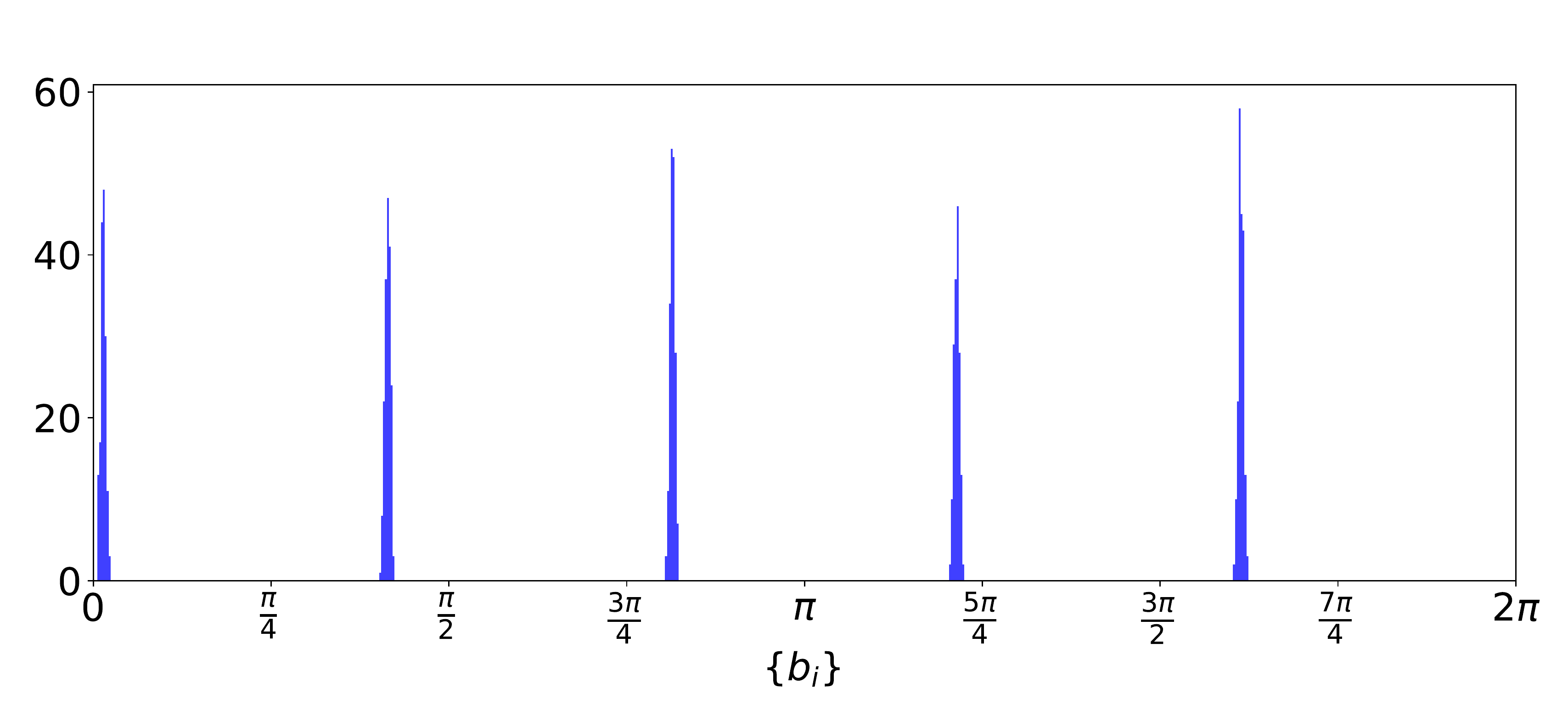}
    \caption{Histogram of extracted gauge variables $\{b_i\}$ for the MXYGG model, revealing five equally-spaced peaks corresponding to the five discrete choices of $b_i$ in the MC simulation. Note that the gauge variables are known up to an overall rotation given by the global $U(1)$ symmetry of the Hamiltonian \eqref{eq:xy-M}.}
    \label{fig:extracted}
\end{figure}

\textit{Results for the MXYGG Model} -- We now turn to the more complex case of the MXYGG model. As in \cite{WangZhaiPCA1,SinghUMLXY}, we first perform PCA on the full dataset $\{\{\cos(\phi_i)\},\{\sin(\phi_i)\}\}$ generated from MC simulations. By projecting this data onto the first two principal components, which are equally most important, the resulting clusters have the same $U(1)$ symmetry as the ones reported for the regular $XY$ model (see Fig. 2 of \cite{WangZhaiPCA1} and discussion therein). This similarity in the clusters suggests that PCA is characterizing the full dataset of the MXYGG and $XY$ models in the same fashion, \textit{i.e.} according to the magnetization vector \cite{WangZhaiPCA1}. However, if PCA is performed only on the $X$ dataset or the $Y$ dataset of the MXYGG model, the resulting clusters \textit{still} reveal a $U(1)$ symmetry, as shown in Fig.~\ref{fig:gauge}; this is in contrast to the results for the regular $XY$ model (see Fig.~\ref{fig:PCA_XY_regular} of the Appendix). This difference indicates that PCA identifies some feature that differentiates the MXYGG model from the regular $XY$ model. This suggests that PCA has detected the Mattis gauge transformation, which must also be present in the full dataset.
 
\indent Now that the presence of the gauge transformation has been identified, we return to the principal components calculated from the full dataset. As in the regular $XY$ model \cite{WangZhaiPCA1}, the first two principal components, $\vec{u}^{(1)}$ and $\vec{u}^{(2)}$, have the largest explained variance ratios. These two eigenvectors describe the non-zero magnetization components observed in the finite system \cite{WangZhaiPCA1}. The projections of the data onto the first and second principal components take the form
\begin{eqnarray}
\begin{split}
\ell^{(1)}&\equiv\sum_i \left( u_i^{(1c)}\cos{(\phi_i)} + u_i^{(1s)}\sin{(\phi_i)}\right) \\
\ell^{(2)}&\equiv\sum_i \left(u_i^{(2c)}\cos{(\phi_i)} + u_i^{(2s)}\sin{(\phi_i)}\right),
\end{split} \label{eq:XYMagProj}
\end{eqnarray}
where we have defined $\vec{u}^{(k)} \equiv (\{u_i^{(kc)}\}, \{u_i^{(ks)}\})$ to match the separation of cosines and sines in the full dataset $\{\{\cos(\phi_i)\},\{\sin(\phi_i)\}\}$. Since we know that PCA is characterizing the data according to the magnetization vector, we identify the projections $\ell^{(1)}$ and $\ell^{(2)}$ with the components of the magnetization in Eq.~\eqref{eq:XYMag}. Through this identification the values of the gauge variables $\{b_i\}$ are extracted from the principal components, as detailed in the Appendix. The distribution of the extracted gauge variables $\{b_i\}$ is shown in Fig.~\ref{fig:extracted}, revealing five equally-spaced peaks as expected for the five equally-spaced choices of gauge variables. PCA is therefore able to calculate the transformation that maps the MXYGG model onto the regular $XY$ model.
\newline

\textit{Conclusion} -- We have applied PCA to two spin models with random interactions, the MISG and MXYGG models on a square lattice. PCA was able to determine that each spin model can be related to a simpler model, namely the regular Ising and $XY$ models. This was accomplished by (1) recognizing the similarities between the projections of the input data onto the principal components of the regular and gauge-transformed models, (2) identifying that PCA characterizes the data using the same thermodynamic quantity (\textit{i.e.} the magnetization), and (3) verifying that the gauge variables calculated by PCA were consistent with the ones selected within MC simulations. These results should easily generalize to other gauge-symmetric spin models with random interactions, such as spin glass models with $O(3)$ gauge symmetry \cite{FischerSpinGlass, NishimoriSpinGlass}.

Our work suggests that UML is capable of more than just classifying data; interpretable UML methods could possibly learn hidden features of an underlying model, such as symmetries and gauge transformations. For the physicist, this means UML could reveal previously unknown insights into a simulated model. It is of interest to investigate how other UML methods beyond PCA fare in this regard (\textit{e.g.} autoencoders, which share some similarities with PCA, are capable of nonlinear fitting and therefore possess greater descriptive power \cite{WetzelAutoencoders}). Such methods may not be as interpretable as PCA; hence, using them to discover the hidden properties of a model might be a more complicated task. However, even in such cases, our work indicates that UML could at least ``see through'' nontrivial characteristics such as gauge symmetries. This suggests that UML methods could alternatively be used to efficiently label data for subsequently applied SML methods, which may explain how PCA and a neural network together learned the $SU(2)$ gauge theory order parameter \cite{WetzelSU2}. The generalization of this idea to other and more powerful UML methods may therefore expedite the learning process of a neural network, which makes this an avenue worth pursuing in its own right and especially for classifying phases. Lastly, gauge-symmetric models represent a class of models with known mathematical simplifications. Applying UML methods to these models may therefore provide a deeper understanding of how these methods work and what exactly they learn. \\

We thank W. Jin, C. X. Cerkauskas, K. Chung, A. Golubeva, R. G. Melko, and S. J. Wetzel for helpful discussions. This work was supported by the Canada Research Chair program (M.J.P.G., Tier 1) and by the NSERC of Canada CGS-M program (D.P.).

\newpage

\section{Appendix}
\subsection{Additional Analysis of the MISG Model}

\subsubsection{Definition of Plaquettes}
\begin{figure}[ht!]
     \centering
     \begin{overpic}[width=\columnwidth]{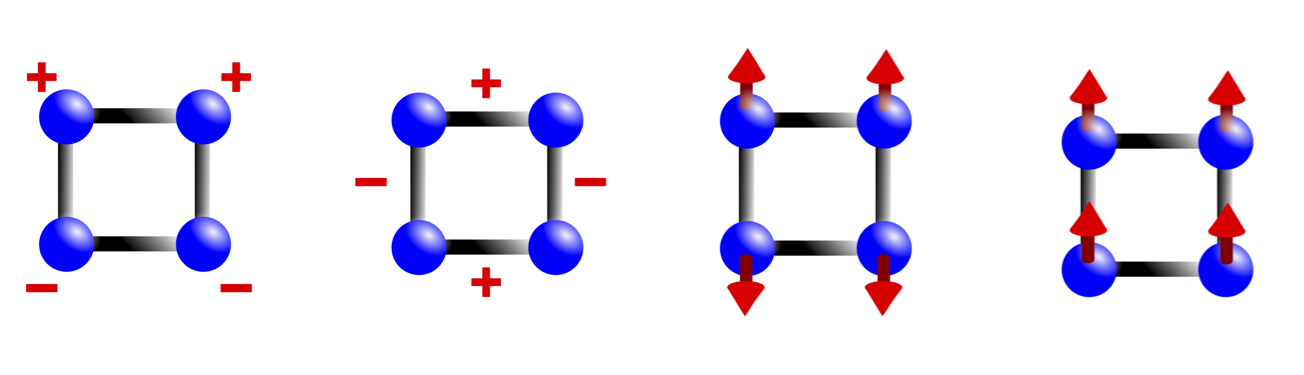}
     \put(8,27){(a)}
     \put(35,27){(b)}
     \put(60,27){(c)}
     \put(86,27){(d)}
    \end{overpic}
     \caption{Plaquette in the MISG model. (a) Example of gauge variables $\{\epsilon_i\}$ for the four sites. (b) Resulting signs of the bond interactions $\{J_{ij}\}$ for the four bonds. (c) Example of ground state spin configuration of $\sigma_i^z$ variables for these random bond interactions. Note that the coupling in the Hamiltonian is $-J_{ij}$. (d) Resulting ground state configuration of $\tau_i^z$ as a product of the spin configuration illustrated in (c) with the gauge variables in (a). }
     \label{fig:Plaquette}
 \end{figure}
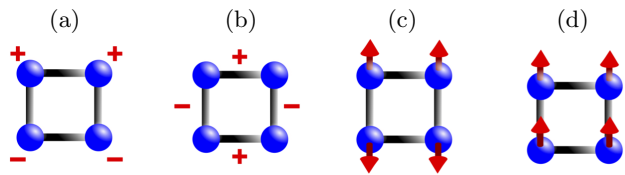

A plaquette in the lattice is defined as the smallest region contained within a closed loop of neighbouring sites. On the square lattice, the resulting plaquettes are composed of four sites. For the Mattis transformation we introduce gauge variables $\epsilon_i$ for every site to define the coupling constant $J_{ij}=\epsilon_i \epsilon_j J$ on every nearest neighbour bond. This procedure is sketched in Fig.~\ref{fig:Plaquette}.

 \subsubsection{MC Simulation with the Learned Gauge Variables}
\begin{figure}
     \centering
     \includegraphics[width=90mm]{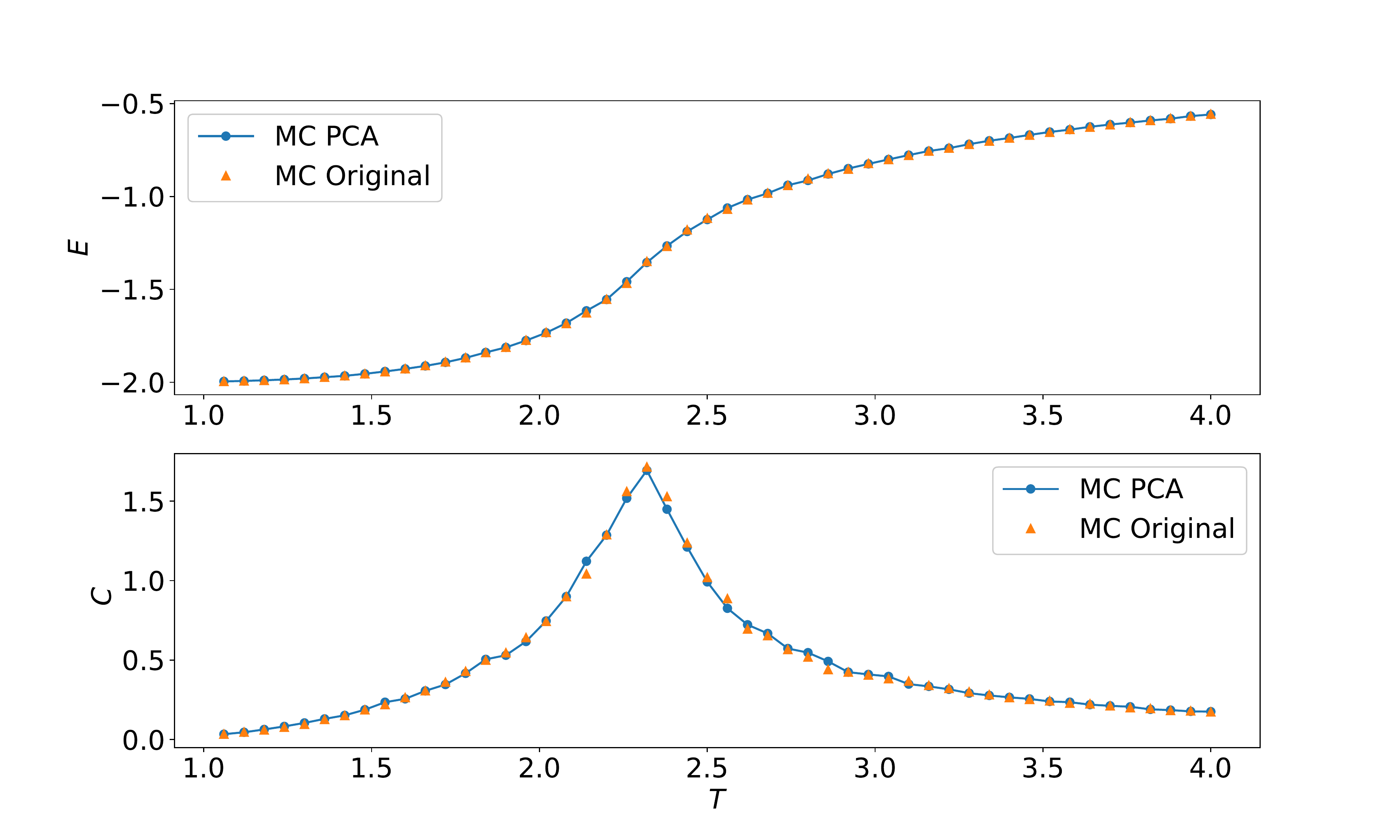}
     \caption{Comparison of thermodynamic quantities (energy per spin $E$ and specific heat $C$) calculated within MC simulations, using the original known gauge variables and the learned gauge variables from PCA.}
     \label{fig:MCRerun}
 \end{figure}

 After applying PCA to the MISG model, we study the faithfulness of the learned gauge variables. We performed a MC simulation on the MISG model as before, but instead used the learned gauge variables in place of the known gauge variables. The thermodynamic quantities obtained with this simulation are then compared with the thermodynamic quantities which used the known gauge variables, as shown in Fig.~\ref{fig:MCRerun}. 
 As can be seen, the energy and specific curves obtained for both simulations are identical, supporting PCA's ability to learn the gauge variables of the MISG model.

\subsection{PCA Results for the Regular Ising and $XY$ Models}

\subsubsection{PCA Clusters for the Regular Ising and $XY$ Models}
\begin{figure}[ht]
     \centering
     \begin{overpic}[width=\columnwidth]{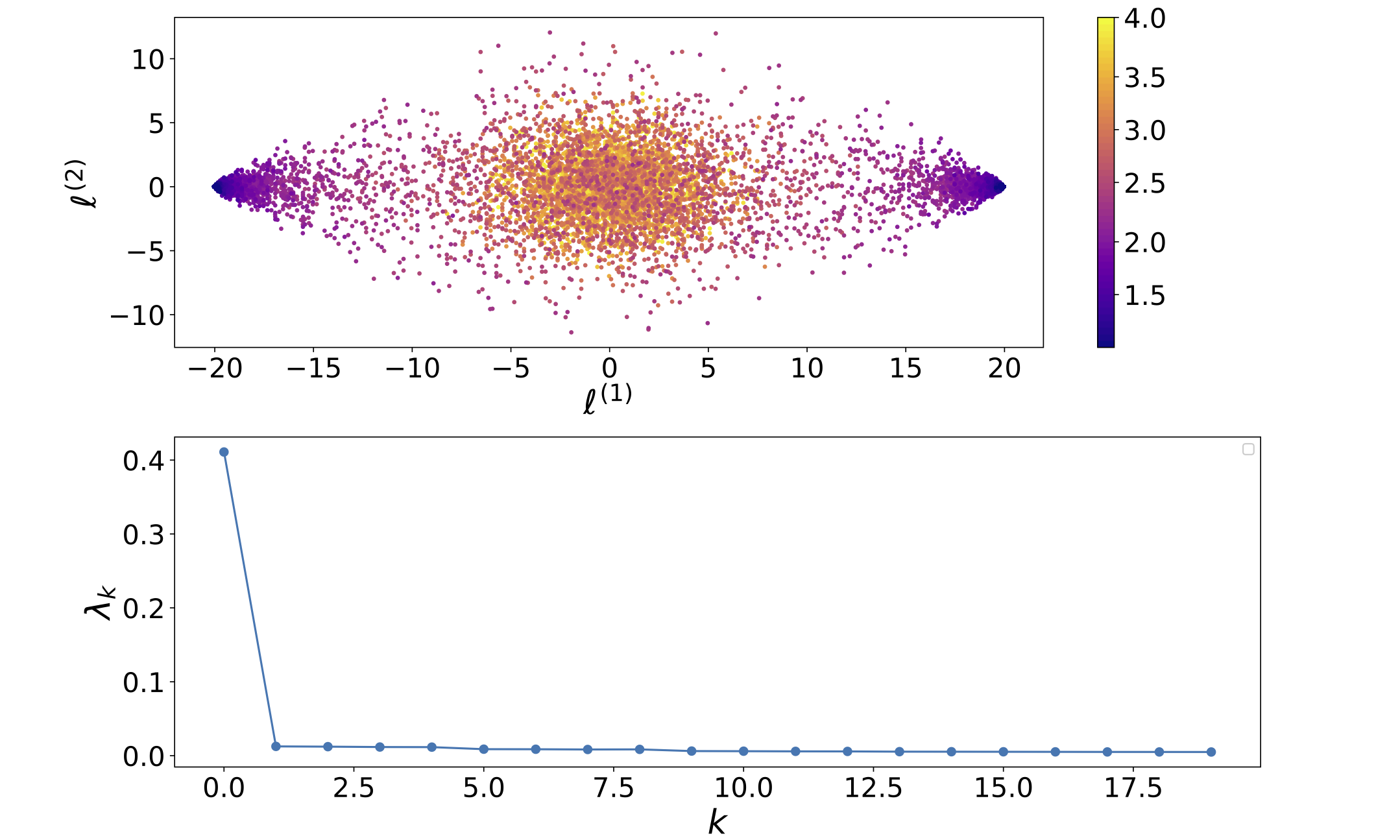}
    \put(68,55){(a)}
    \put(84,25){(b)}
    \end{overpic}
     \caption{(a) Principal component projection $\ell^{(1)}$ versus $\ell^{(2)}$ and (b) first 20 explained variance ratios for the regular Ising model on an $L$ = 20 square lattice.}
     \label{fig:PCA_Ising}
 \end{figure}

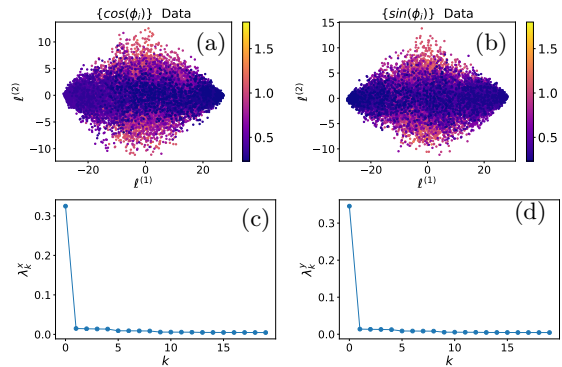
\begin{figure}[ht]
    \centering
    \begin{overpic}[width=\columnwidth]{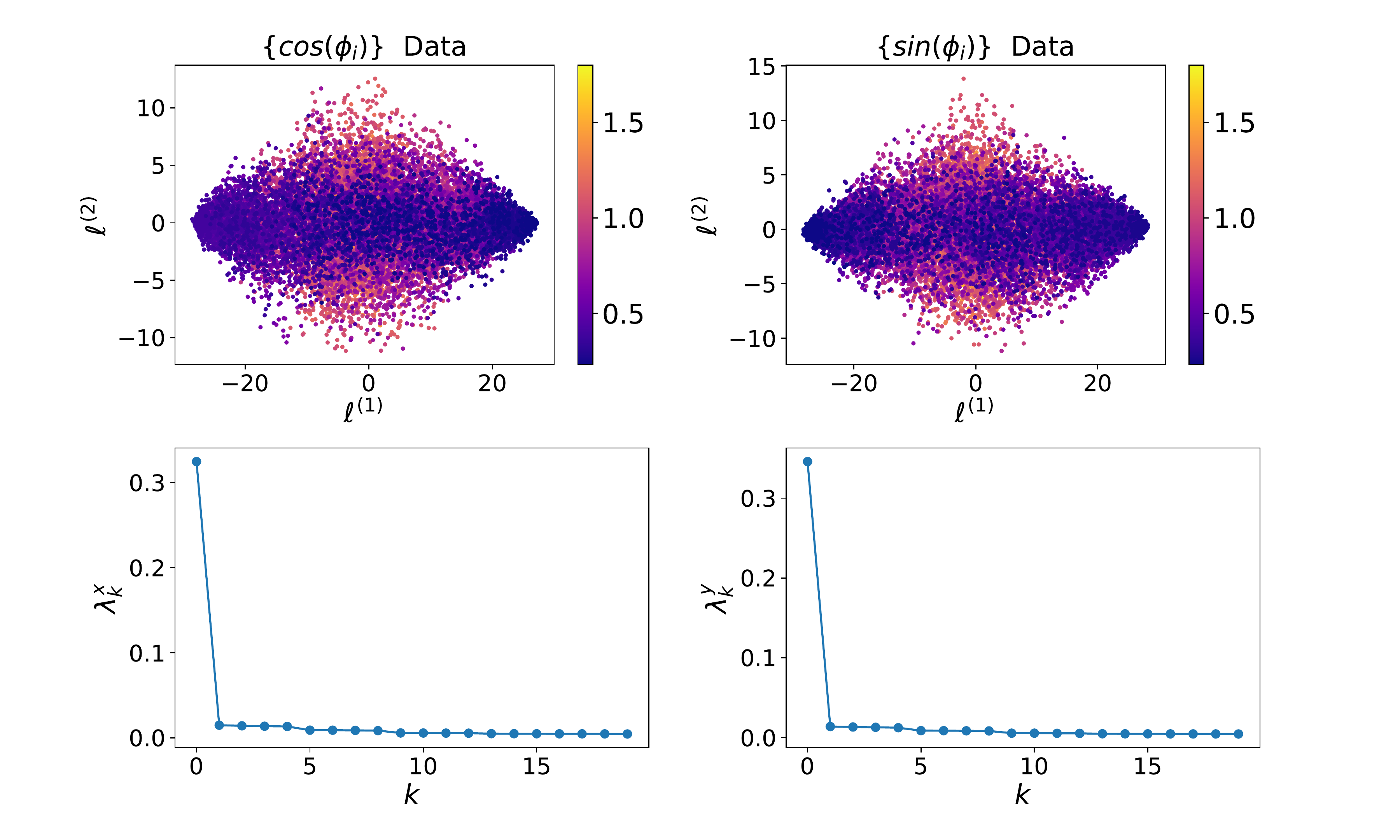}
    \put(34,51){(a)}
    \put(77,51){(b)}
    \put(41,24){(c)}
    \put(83,25){(d)}
    \end{overpic}
    \caption{Principal component projections $\ell^{(1)}$ versus $\ell^{(2)}$ ((a) and (b)) and first 20 explained variance ratios ((c) and (d)) for the regular $XY$ model, for PCA applied to $\{\cos{(\phi_i)}\}$ or $\{\sin{(\phi_i)}\}$ only. The configurations were sampled from an $L = 30$ square lattice.}
    \label{fig:PCA_XY_regular}
\end{figure}

For completeness and comparison, MC simulations are run for the regular Ising model ($L$ = 20) and the regular $XY$ model ($L$ = 30) on a square lattice; the exact same parameters for the MC simulation temperature sweeps as detailed in the main text for the MISG and MXYGG models are used here. PCA is applied to the spin configurations from both sets of data, which are formatted in the same manner as the input data of the MISG and MXYGG models. The clusters identified by PCA for the regular Ising model are shown in Fig.~\ref{fig:PCA_Ising}. For the regular $XY$ model, PCA is applied to either the $X$ dataset ($\{\cos(\phi_i)\}$) or the $Y$ dataset ($\{\sin(\phi_i)\}$). The projections onto the first two principal components of the $X$ dataset alone or the $Y$ dataset alone are shown in Fig.~\ref{fig:PCA_XY_regular}. Firstly, Fig.~\ref{fig:PCA_XY_regular} should be compared with Fig.~\ref{fig:gauge} of the main text. Although the clusters that PCA identifies for the regular $XY$ and the MXYGG models look the same when provided with the \textit{full} dataset, there is a clear difference when PCA is provided with only the $X$ or $Y$ dataset. This difference is indicative of an identified feature which is not present in the regular $XY$ model. Secondly, Fig.~\ref{fig:PCA_XY_regular} should be compared with Fig.~\ref{fig:PCA_Ising}. Previous work on the regular Ising model \cite{WangUnsupervisedIsing} has shown that the central high-temperature cluster and the two adjacent low-temperature clusters correspond to the paramagnetic and ferromagnetic phases, respectively, which PCA determines by summing the spin configurations. Fig.~\ref{fig:PCA_XY_regular} can be similarly interpreted in light of this. PCA characterizes the input data of the regular $XY$ model by directly summing the spin configurations along two orthogonal directions. This explains why the PCA clusters for the $X$ and $Y$ datasets of the regular $XY$ model look like those of the regular Ising model. However, since the spin variables in the regular $XY$ model are continuous and not discrete $\pm1$ values as in the regular Ising model, the low-temperature projection forms one continuous line rather than two separate clusters. Note that the two orthogonal directions along which the magnetization is determined by PCA are not necessarily the chosen $x$ and $y$ directions of the MC simulation, owing to the global $U(1)$ rotational symmetry of the regular $XY$ model. The determination of this global rotation is the focus of the next section.

\subsubsection{Proof of Global Rotation for the Regular $XY$ Model}

\begin{figure}
\centering
    \begin{overpic}[width=\columnwidth]{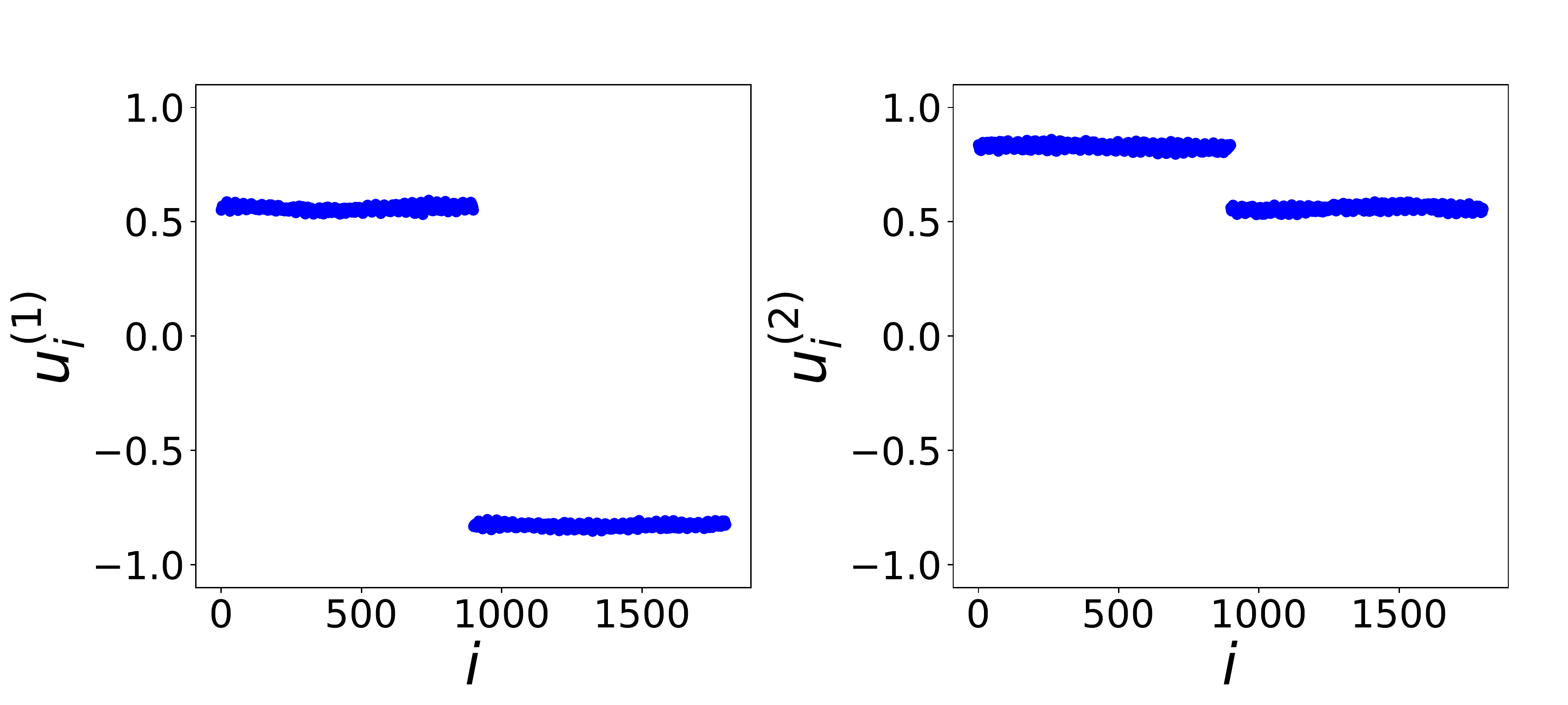}
    \put(25,42){(a)}
    \put(77,42){(b)}
    \end{overpic}
    \caption{Components of the (a) first and (b) second principal component eigenvectors, $\vec{u}^{(1)}$ and $\vec{u}^{(2)}$, for the regular $XY$ model before applying a global rotation. The two branches of each graph correspond to coefficients for $\{\cos{(\phi_i)}\}$ or $\{\sin{(\phi_i)}\}$ data.}
    \label{fig:Prerotation}
\end{figure}

\begin{figure}
\centering
    \includegraphics[width=\columnwidth]{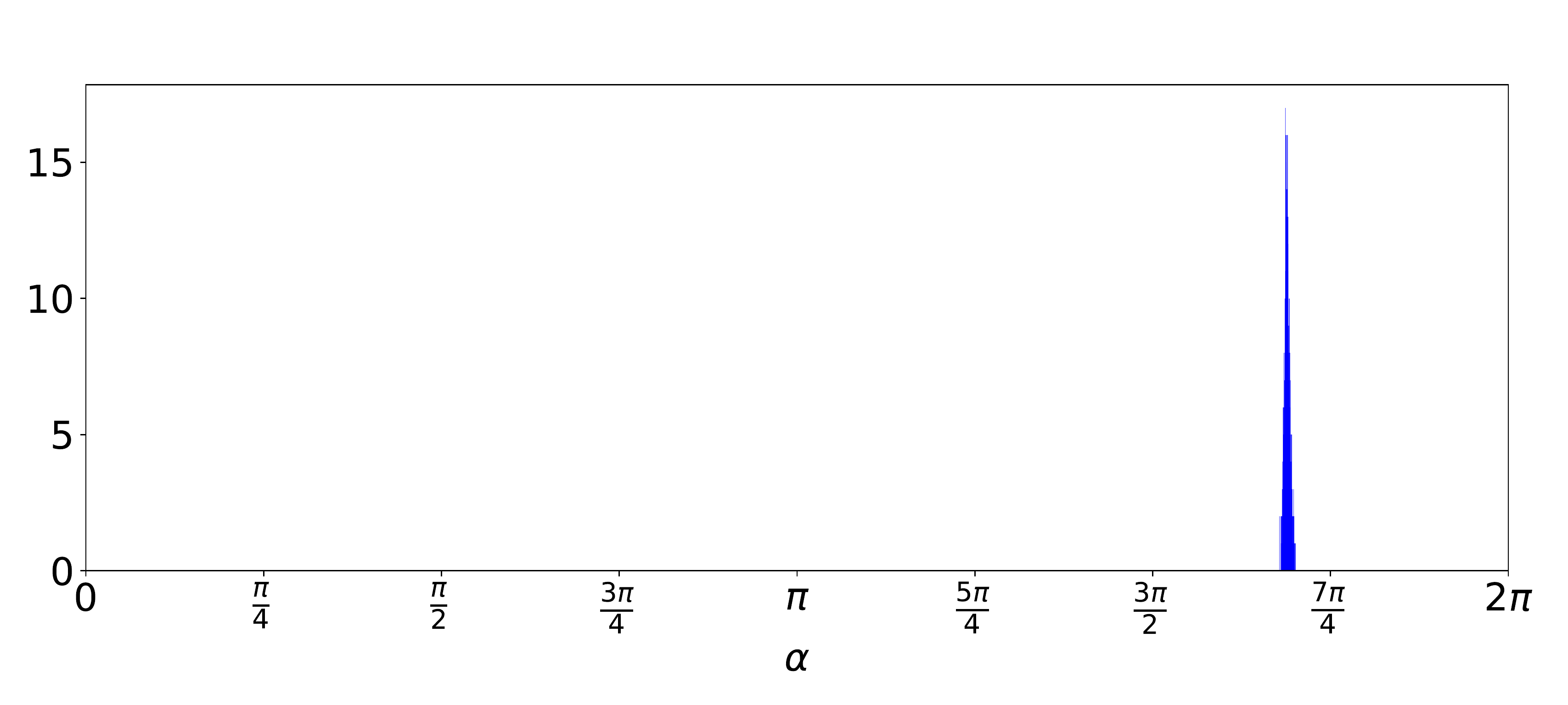}
    \caption{Histogram of extracted global rotation angle $\alpha$ for the regular $XY$ model.}
    \label{fig:extracted_theta}
\end{figure}

\begin{figure}
\centering
    \begin{overpic}[width=\columnwidth]{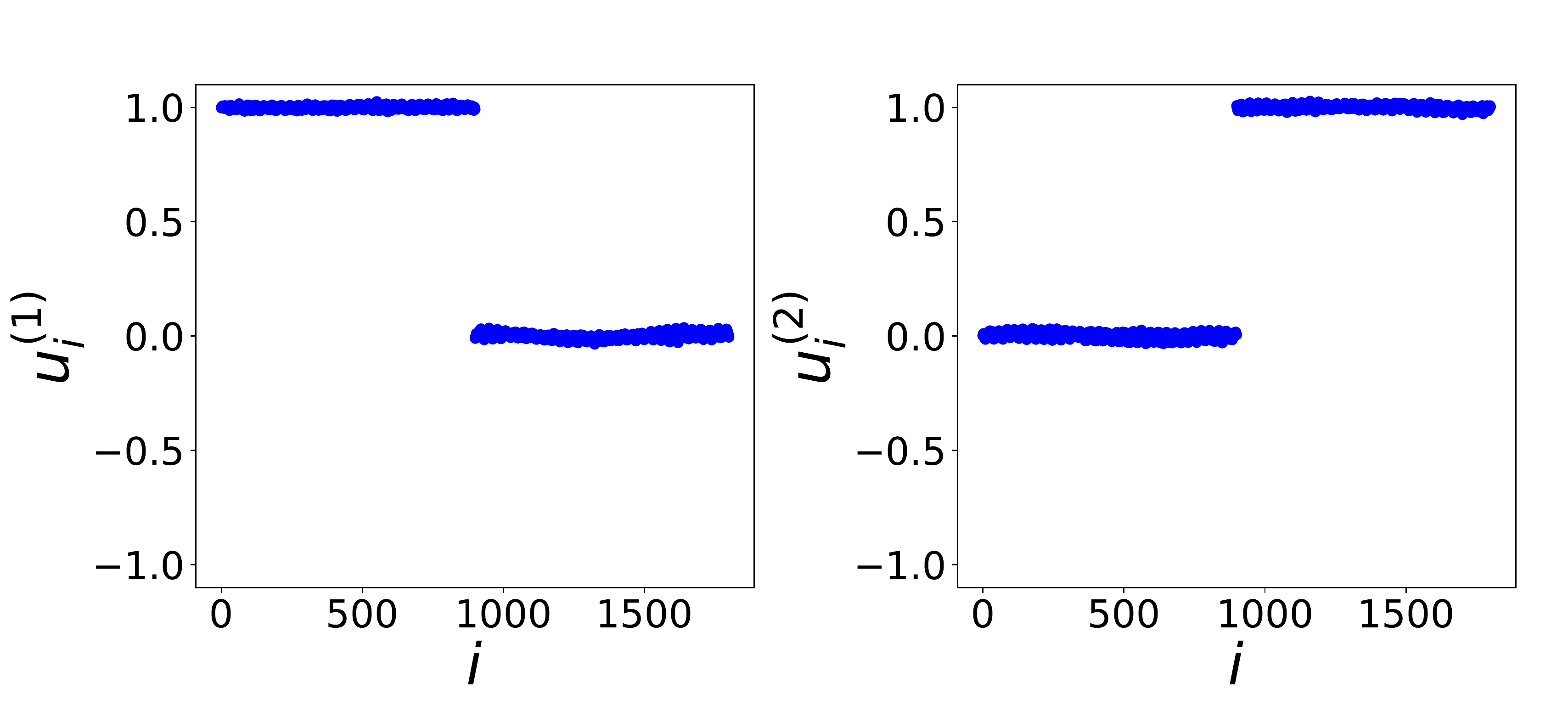}
    \put(25,42){(a)}
    \put(77,42){(b)}
    \end{overpic}
    \caption{Components of the (a) first and (b) second principal component eigenvectors, $\vec{u}^{(1)}$ and $\vec{u}^{(2)}$, for the regular $XY$ model after applying a global rotation. The two branches of each graph correspond to coefficients for $\{\cos{(\phi_i)}\}$ or $\{\sin{(\phi_i)}\}$ data.}
    \label{fig:Postrotation}
\end{figure}

The regular $XY$ model can be considered as the MXYGG model with $b_i = 0$ for all lattice sites $i$. In this case, the magnetization vector for the regular $XY$ model is
\begin{eqnarray}
M_x&=&\sum_i \cos{(\phi_i)}\nonumber\\
M_y&=&\sum_i \sin{(\phi_i)}\nonumber,
\end{eqnarray}
in contrast to Eq.~\eqref{eq:XYMag}. Comparing with the principal component projections in Eq.~\eqref{eq:XYMagProj}, this would imply that the principal component eigenvectors should have components $\{u_i^{1c}\} = \{u_i^{2s}\} = 1$ and $\{u_i^{1s}\} = \{u_i^{2c}\} = 0$ if PCA was learning the magnetization along the $x$ and $y$ directions from MC simulations. This is clearly not the case; see Fig.~\ref{fig:Prerotation}. However, accounting for the global $U(1)$ rotation symmetry of the $XY$ model, the magnetization takes the general form
\begin{eqnarray}
M_x&=&\sum_i \cos(\phi_i + \alpha)= \sum_i\left( \cos{\phi_i}\cos{\alpha} - \sin{\phi_i}\sin{\alpha}\right),\nonumber\\
M_y&=&\sum_i \sin(\phi_i + \alpha)= \sum_i\left( \sin{\phi_i}\cos{\alpha} + \cos{\phi_i}\sin{\alpha}\right)\nonumber,
\end{eqnarray}
for a global rotation angle $\alpha$. Comparing this expression with Eq.~\eqref{eq:XYMagProj}, the components of the principal component eigenvectors in Fig.~\ref{fig:Prerotation} therefore indicate the global rotation $\alpha$ along which PCA learns the magnetization. By considering this global rotation, the components of these principal component eigenvectors can be used to analytically determine the value of $\alpha$; the histogram for this extraction is shown in Fig.~\ref{fig:extracted_theta}. When this global rotation is accounted for, the principal eigenvectors do take values of only $1$s or $0$s, as shown in Fig.~\ref{fig:Postrotation}. This proves that PCA is learning the magnetization of the $XY$ model along two orthogonal directions; this same analysis can be applied to the principal components of the MXYGG model to extract the \textit{local} rotations produced by the gauge variables $\{b_i\}$, giving the histogram in Fig.~\ref{fig:extracted}.

\bibliography{refs}

\begin{thebibliography}{40}%
\makeatletter
\providecommand \@ifxundefined [1]{%
 \@ifx{#1\undefined}
}%
\providecommand \@ifnum [1]{%
 \ifnum #1\expandafter \@firstoftwo
 \else \expandafter \@secondoftwo
 \fi
}%
\providecommand \@ifx [1]{%
 \ifx #1\expandafter \@firstoftwo
 \else \expandafter \@secondoftwo
 \fi
}%
\providecommand \natexlab [1]{#1}%
\providecommand \enquote  [1]{``#1''}%
\providecommand \bibnamefont  [1]{#1}%
\providecommand \bibfnamefont [1]{#1}%
\providecommand \citenamefont [1]{#1}%
\providecommand \href@noop [0]{\@secondoftwo}%
\providecommand \href [0]{\begingroup \@sanitize@url \@href}%
\providecommand \@href[1]{\@@startlink{#1}\@@href}%
\providecommand \@@href[1]{\endgroup#1\@@endlink}%
\providecommand \@sanitize@url [0]{\catcode `\\12\catcode `\$12\catcode
  `\&12\catcode `\#12\catcode `\^12\catcode `\_12\catcode `\%12\relax}%
\providecommand \@@startlink[1]{}%
\providecommand \@@endlink[0]{}%
\providecommand \url  [0]{\begingroup\@sanitize@url \@url }%
\providecommand \@url [1]{\endgroup\@href {#1}{\urlprefix }}%
\providecommand \urlprefix  [0]{URL }%
\providecommand \Eprint [0]{\href }%
\providecommand \doibase [0]{http://dx.doi.org/}%
\providecommand \selectlanguage [0]{\@gobble}%
\providecommand \bibinfo  [0]{\@secondoftwo}%
\providecommand \bibfield  [0]{\@secondoftwo}%
\providecommand \translation [1]{[#1]}%
\providecommand \BibitemOpen [0]{}%
\providecommand \bibitemStop [0]{}%
\providecommand \bibitemNoStop [0]{.\EOS\space}%
\providecommand \EOS [0]{\spacefactor3000\relax}%
\providecommand \BibitemShut  [1]{\csname bibitem#1\endcsname}%
\let\auto@bib@innerbib\@empty
\bibitem [{\citenamefont {Wang}(2016)}]{WangUnsupervisedIsing}%
  \BibitemOpen
  \bibfield  {author} {\bibinfo {author} {\bibfnamefont {L.}~\bibnamefont
  {Wang}},\ }\href {\doibase 10.1103/PhysRevB.94.195105} {\bibfield  {journal}
  {\bibinfo  {journal} {Phys. Rev. B}\ }\textbf {\bibinfo {volume} {94}},\
  \bibinfo {pages} {195105} (\bibinfo {year} {2016})}\BibitemShut {NoStop}%
\bibitem [{\citenamefont {Carrasquilla}\ and\ \citenamefont
  {Melko}(2017)}]{MelkoIsingNN}%
  \BibitemOpen
  \bibfield  {author} {\bibinfo {author} {\bibfnamefont {J.}~\bibnamefont
  {Carrasquilla}}\ and\ \bibinfo {author} {\bibfnamefont {R.~G.}\ \bibnamefont
  {Melko}},\ }\href {\doibase 10.1038/nphys4035} {\bibfield  {journal}
  {\bibinfo  {journal} {Nat. Phys}\ }\textbf {\bibinfo {volume} {13}},\
  \bibinfo {pages} {431} (\bibinfo {year} {2017})}\BibitemShut {NoStop}%
\bibitem [{\citenamefont {{Mehta}}\ \emph {et~al.}(2019)\citenamefont
  {{Mehta}}, \citenamefont {{Bukov}}, \citenamefont {{Wang}}, \citenamefont
  {{Day}}, \citenamefont {{Richardson}}, \citenamefont {{Fisher}},\ and\
  \citenamefont {{Schwab}}}]{MehtaRev}%
  \BibitemOpen
  \bibfield  {author} {\bibinfo {author} {\bibfnamefont {P.}~\bibnamefont
  {{Mehta}}}, \bibinfo {author} {\bibfnamefont {M.}~\bibnamefont {{Bukov}}},
  \bibinfo {author} {\bibfnamefont {C.-H.}\ \bibnamefont {{Wang}}}, \bibinfo
  {author} {\bibfnamefont {A.~r. G.~R.}\ \bibnamefont {{Day}}}, \bibinfo
  {author} {\bibfnamefont {C.}~\bibnamefont {{Richardson}}}, \bibinfo {author}
  {\bibfnamefont {C.~K.}\ \bibnamefont {{Fisher}}}, \ and\ \bibinfo {author}
  {\bibfnamefont {D.~J.}\ \bibnamefont {{Schwab}}},\ }\href {\doibase
  10.1016/j.physrep.2019.03.001} {\bibfield  {journal} {\bibinfo  {journal}
  {Phys. Rep.}\ }\textbf {\bibinfo {volume} {810}},\ \bibinfo {pages} {1}
  (\bibinfo {year} {2019})}\BibitemShut {NoStop}%
\bibitem [{\citenamefont {Dunjko}\ and\ \citenamefont
  {Briegel}(2018)}]{DunjkoRev}%
  \BibitemOpen
  \bibfield  {author} {\bibinfo {author} {\bibfnamefont {V.}~\bibnamefont
  {Dunjko}}\ and\ \bibinfo {author} {\bibfnamefont {H.~J.}\ \bibnamefont
  {Briegel}},\ }\href {\doibase 10.1088/1361-6633/aab406} {\bibfield  {journal}
  {\bibinfo  {journal} {Reports on Progress in Physics}\ }\textbf {\bibinfo
  {volume} {81}},\ \bibinfo {pages} {074001} (\bibinfo {year}
  {2018})}\BibitemShut {NoStop}%
\bibitem [{\citenamefont {Carleo}\ \emph {et~al.}(2019)\citenamefont {Carleo},
  \citenamefont {Cirac}, \citenamefont {Cranmer}, \citenamefont {Daudet},
  \citenamefont {Schuld}, \citenamefont {Tishby}, \citenamefont
  {Vogt-Maranto},\ and\ \citenamefont {Zdeborov\'a}}]{MLinPhysicsReview2019}%
  \BibitemOpen
  \bibfield  {author} {\bibinfo {author} {\bibfnamefont {G.}~\bibnamefont
  {Carleo}}, \bibinfo {author} {\bibfnamefont {I.}~\bibnamefont {Cirac}},
  \bibinfo {author} {\bibfnamefont {K.}~\bibnamefont {Cranmer}}, \bibinfo
  {author} {\bibfnamefont {L.}~\bibnamefont {Daudet}}, \bibinfo {author}
  {\bibfnamefont {M.}~\bibnamefont {Schuld}}, \bibinfo {author} {\bibfnamefont
  {N.}~\bibnamefont {Tishby}}, \bibinfo {author} {\bibfnamefont
  {L.}~\bibnamefont {Vogt-Maranto}}, \ and\ \bibinfo {author} {\bibfnamefont
  {L.}~\bibnamefont {Zdeborov\'a}},\ }\href {\doibase
  10.1103/RevModPhys.91.045002} {\bibfield  {journal} {\bibinfo  {journal}
  {Rev. Mod. Phys.}\ }\textbf {\bibinfo {volume} {91}},\ \bibinfo {pages}
  {045002} (\bibinfo {year} {2019})}\BibitemShut {NoStop}%
\bibitem [{\citenamefont {Zhao}\ \emph {et~al.}(2019)\citenamefont {Zhao},
  \citenamefont {Kao}, \citenamefont {Wu},\ and\ \citenamefont
  {Kao}}]{KaoReinforceMC}%
  \BibitemOpen
  \bibfield  {author} {\bibinfo {author} {\bibfnamefont {K.-W.}\ \bibnamefont
  {Zhao}}, \bibinfo {author} {\bibfnamefont {W.-H.}\ \bibnamefont {Kao}},
  \bibinfo {author} {\bibfnamefont {K.-H.}\ \bibnamefont {Wu}}, \ and\ \bibinfo
  {author} {\bibfnamefont {Y.-J.}\ \bibnamefont {Kao}},\ }\href {\doibase
  10.1103/PhysRevE.99.062106} {\bibfield  {journal} {\bibinfo  {journal} {Phys.
  Rev. E}\ }\textbf {\bibinfo {volume} {99}},\ \bibinfo {pages} {062106}
  (\bibinfo {year} {2019})}\BibitemShut {NoStop}%
\bibitem [{\citenamefont {Greitemann}\ \emph
  {et~al.}(2019{\natexlab{a}})\citenamefont {Greitemann}, \citenamefont {Liu},\
  and\ \citenamefont {Pollet}}]{GreitemannSVM}%
  \BibitemOpen
  \bibfield  {author} {\bibinfo {author} {\bibfnamefont {J.}~\bibnamefont
  {Greitemann}}, \bibinfo {author} {\bibfnamefont {K.}~\bibnamefont {Liu}}, \
  and\ \bibinfo {author} {\bibfnamefont {L.}~\bibnamefont {Pollet}},\ }\href
  {\doibase 10.1103/PhysRevB.99.060404} {\bibfield  {journal} {\bibinfo
  {journal} {Phys. Rev. B}\ }\textbf {\bibinfo {volume} {99}},\ \bibinfo
  {pages} {060404} (\bibinfo {year} {2019}{\natexlab{a}})}\BibitemShut
  {NoStop}%
\bibitem [{\citenamefont {Beach}\ \emph {et~al.}(2018)\citenamefont {Beach},
  \citenamefont {Golubeva},\ and\ \citenamefont {Melko}}]{BeachMLVortices}%
  \BibitemOpen
  \bibfield  {author} {\bibinfo {author} {\bibfnamefont {M.~J.~S.}\
  \bibnamefont {Beach}}, \bibinfo {author} {\bibfnamefont {A.}~\bibnamefont
  {Golubeva}}, \ and\ \bibinfo {author} {\bibfnamefont {R.~G.}\ \bibnamefont
  {Melko}},\ }\href {\doibase 10.1103/PhysRevB.97.045207} {\bibfield  {journal}
  {\bibinfo  {journal} {Phys. Rev. B}\ }\textbf {\bibinfo {volume} {97}},\
  \bibinfo {pages} {045207} (\bibinfo {year} {2018})}\BibitemShut {NoStop}%
\bibitem [{\citenamefont {Greitemann}\ \emph
  {et~al.}(2019{\natexlab{b}})\citenamefont {Greitemann}, \citenamefont {Liu},
  \citenamefont {Jaubert}, \citenamefont {Yan}, \citenamefont {Shannon},\ and\
  \citenamefont {Pollet}}]{GreitemannHiddenOrder}%
  \BibitemOpen
  \bibfield  {author} {\bibinfo {author} {\bibfnamefont {J.}~\bibnamefont
  {Greitemann}}, \bibinfo {author} {\bibfnamefont {K.}~\bibnamefont {Liu}},
  \bibinfo {author} {\bibfnamefont {L.~D.~C.}\ \bibnamefont {Jaubert}},
  \bibinfo {author} {\bibfnamefont {H.}~\bibnamefont {Yan}}, \bibinfo {author}
  {\bibfnamefont {N.}~\bibnamefont {Shannon}}, \ and\ \bibinfo {author}
  {\bibfnamefont {L.}~\bibnamefont {Pollet}},\ }\href {\doibase
  10.1103/PhysRevB.100.174408} {\bibfield  {journal} {\bibinfo  {journal}
  {Phys. Rev. B}\ }\textbf {\bibinfo {volume} {100}},\ \bibinfo {pages}
  {174408} (\bibinfo {year} {2019}{\natexlab{b}})}\BibitemShut {NoStop}%
\bibitem [{\citenamefont {Ponte}\ and\ \citenamefont
  {Melko}(2017{\natexlab{a}})}]{Ponte_kernel}%
  \BibitemOpen
  \bibfield  {author} {\bibinfo {author} {\bibfnamefont {P.}~\bibnamefont
  {Ponte}}\ and\ \bibinfo {author} {\bibfnamefont {R.~G.}\ \bibnamefont
  {Melko}},\ }\href {\doibase 10.1103/PhysRevB.96.205146} {\bibfield  {journal}
  {\bibinfo  {journal} {Phys. Rev. B}\ }\textbf {\bibinfo {volume} {96}},\
  \bibinfo {pages} {205146} (\bibinfo {year} {2017}{\natexlab{a}})}\BibitemShut
  {NoStop}%
\bibitem [{\citenamefont {Liu}\ \emph {et~al.}(2019)\citenamefont {Liu},
  \citenamefont {Greitemann},\ and\ \citenamefont {Pollet}}]{Pollet_svm}%
  \BibitemOpen
  \bibfield  {author} {\bibinfo {author} {\bibfnamefont {K.}~\bibnamefont
  {Liu}}, \bibinfo {author} {\bibfnamefont {J.}~\bibnamefont {Greitemann}}, \
  and\ \bibinfo {author} {\bibfnamefont {L.}~\bibnamefont {Pollet}},\ }\href
  {\doibase 10.1103/PhysRevB.99.104410} {\bibfield  {journal} {\bibinfo
  {journal} {Phys. Rev. B}\ }\textbf {\bibinfo {volume} {99}},\ \bibinfo
  {pages} {104410} (\bibinfo {year} {2019})}\BibitemShut {NoStop}%
\bibitem [{\citenamefont {Th\'eveniaut}\ and\ \citenamefont
  {Alet}(2019)}]{PhysRevB.100.224202}%
  \BibitemOpen
  \bibfield  {author} {\bibinfo {author} {\bibfnamefont {H.}~\bibnamefont
  {Th\'eveniaut}}\ and\ \bibinfo {author} {\bibfnamefont {F.}~\bibnamefont
  {Alet}},\ }\href {\doibase 10.1103/PhysRevB.100.224202} {\bibfield  {journal}
  {\bibinfo  {journal} {Phys. Rev. B}\ }\textbf {\bibinfo {volume} {100}},\
  \bibinfo {pages} {224202} (\bibinfo {year} {2019})}\BibitemShut {NoStop}%
\bibitem [{\citenamefont {Canabarro}\ \emph
  {et~al.}(2019{\natexlab{a}})\citenamefont {Canabarro}, \citenamefont
  {Brito},\ and\ \citenamefont {Chaves}}]{PhysRevLett.122.200401}%
  \BibitemOpen
  \bibfield  {author} {\bibinfo {author} {\bibfnamefont {A.}~\bibnamefont
  {Canabarro}}, \bibinfo {author} {\bibfnamefont {S.}~\bibnamefont {Brito}}, \
  and\ \bibinfo {author} {\bibfnamefont {R.}~\bibnamefont {Chaves}},\ }\href
  {\doibase 10.1103/PhysRevLett.122.200401} {\bibfield  {journal} {\bibinfo
  {journal} {Phys. Rev. Lett.}\ }\textbf {\bibinfo {volume} {122}},\ \bibinfo
  {pages} {200401} (\bibinfo {year} {2019}{\natexlab{a}})}\BibitemShut
  {NoStop}%
\bibitem [{\citenamefont {Liang}\ \emph {et~al.}(2018)\citenamefont {Liang},
  \citenamefont {Liu}, \citenamefont {Lin}, \citenamefont {Guo}, \citenamefont
  {Zhang},\ and\ \citenamefont {He}}]{PhysRevB.98.104426}%
  \BibitemOpen
  \bibfield  {author} {\bibinfo {author} {\bibfnamefont {X.}~\bibnamefont
  {Liang}}, \bibinfo {author} {\bibfnamefont {W.-Y.}\ \bibnamefont {Liu}},
  \bibinfo {author} {\bibfnamefont {P.-Z.}\ \bibnamefont {Lin}}, \bibinfo
  {author} {\bibfnamefont {G.-C.}\ \bibnamefont {Guo}}, \bibinfo {author}
  {\bibfnamefont {Y.-S.}\ \bibnamefont {Zhang}}, \ and\ \bibinfo {author}
  {\bibfnamefont {L.}~\bibnamefont {He}},\ }\href {\doibase
  10.1103/PhysRevB.98.104426} {\bibfield  {journal} {\bibinfo  {journal} {Phys.
  Rev. B}\ }\textbf {\bibinfo {volume} {98}},\ \bibinfo {pages} {104426}
  (\bibinfo {year} {2018})}\BibitemShut {NoStop}%
\bibitem [{\citenamefont {August}\ and\ \citenamefont
  {Ni}(2017)}]{PhysRevA.95.012335}%
  \BibitemOpen
  \bibfield  {author} {\bibinfo {author} {\bibfnamefont {M.}~\bibnamefont
  {August}}\ and\ \bibinfo {author} {\bibfnamefont {X.}~\bibnamefont {Ni}},\
  }\href {\doibase 10.1103/PhysRevA.95.012335} {\bibfield  {journal} {\bibinfo
  {journal} {Phys. Rev. A}\ }\textbf {\bibinfo {volume} {95}},\ \bibinfo
  {pages} {012335} (\bibinfo {year} {2017})}\BibitemShut {NoStop}%
\bibitem [{\citenamefont {Decelle}\ \emph {et~al.}(2019)\citenamefont
  {Decelle}, \citenamefont {Martin-Mayor},\ and\ \citenamefont
  {Seoane}}]{PhysRevE.100.050102}%
  \BibitemOpen
  \bibfield  {author} {\bibinfo {author} {\bibfnamefont {A.}~\bibnamefont
  {Decelle}}, \bibinfo {author} {\bibfnamefont {V.}~\bibnamefont
  {Martin-Mayor}}, \ and\ \bibinfo {author} {\bibfnamefont {B.}~\bibnamefont
  {Seoane}},\ }\href {\doibase 10.1103/PhysRevE.100.050102} {\bibfield
  {journal} {\bibinfo  {journal} {Phys. Rev. E}\ }\textbf {\bibinfo {volume}
  {100}},\ \bibinfo {pages} {050102} (\bibinfo {year} {2019})}\BibitemShut
  {NoStop}%
\bibitem [{\citenamefont {{Xu}}\ \emph {et~al.}(2019)\citenamefont {{Xu}},
  \citenamefont {{Li}}, \citenamefont {{Liu}}, \citenamefont {{Wang}},
  \citenamefont {{Yuan}},\ and\ \citenamefont {{Wang}}}]{2019npjQI}%
  \BibitemOpen
  \bibfield  {author} {\bibinfo {author} {\bibfnamefont {H.}~\bibnamefont
  {{Xu}}}, \bibinfo {author} {\bibfnamefont {J.}~\bibnamefont {{Li}}}, \bibinfo
  {author} {\bibfnamefont {L.}~\bibnamefont {{Liu}}}, \bibinfo {author}
  {\bibfnamefont {Y.}~\bibnamefont {{Wang}}}, \bibinfo {author} {\bibfnamefont
  {H.}~\bibnamefont {{Yuan}}}, \ and\ \bibinfo {author} {\bibfnamefont
  {X.}~\bibnamefont {{Wang}}},\ }\href {\doibase 10.1038/s41534-019-0198-z}
  {\bibfield  {journal} {\bibinfo  {journal} {npj Quantum Inf.}\ }\textbf
  {\bibinfo {volume} {5}},\ \bibinfo {eid} {82} (\bibinfo {year}
  {2019})}\BibitemShut {NoStop}%
\bibitem [{\citenamefont {{Bohrdt}}\ \emph {et~al.}(2019)\citenamefont
  {{Bohrdt}}, \citenamefont {{Chiu}}, \citenamefont {{Ji}}, \citenamefont
  {{Xu}}, \citenamefont {{Greif}}, \citenamefont {{Greiner}}, \citenamefont
  {{Demler}}, \citenamefont {{Grusdt}},\ and\ \citenamefont
  {{Knap}}}]{2019NatPh15}%
  \BibitemOpen
  \bibfield  {author} {\bibinfo {author} {\bibfnamefont {A.}~\bibnamefont
  {{Bohrdt}}}, \bibinfo {author} {\bibfnamefont {C.~S.}\ \bibnamefont
  {{Chiu}}}, \bibinfo {author} {\bibfnamefont {G.}~\bibnamefont {{Ji}}},
  \bibinfo {author} {\bibfnamefont {M.}~\bibnamefont {{Xu}}}, \bibinfo {author}
  {\bibfnamefont {D.}~\bibnamefont {{Greif}}}, \bibinfo {author} {\bibfnamefont
  {M.}~\bibnamefont {{Greiner}}}, \bibinfo {author} {\bibfnamefont
  {E.}~\bibnamefont {{Demler}}}, \bibinfo {author} {\bibfnamefont
  {F.}~\bibnamefont {{Grusdt}}}, \ and\ \bibinfo {author} {\bibfnamefont
  {M.}~\bibnamefont {{Knap}}},\ }\href {\doibase 10.1038/s41567-019-0565-x}
  {\bibfield  {journal} {\bibinfo  {journal} {Nat. Phys}\ }\textbf {\bibinfo
  {volume} {15}},\ \bibinfo {pages} {921} (\bibinfo {year} {2019})}\BibitemShut
  {NoStop}%
\bibitem [{\citenamefont {Casert}\ \emph {et~al.}(2019)\citenamefont {Casert},
  \citenamefont {Vieijra}, \citenamefont {Nys},\ and\ \citenamefont
  {Ryckebusch}}]{PhysRevE.99.023304}%
  \BibitemOpen
  \bibfield  {author} {\bibinfo {author} {\bibfnamefont {C.}~\bibnamefont
  {Casert}}, \bibinfo {author} {\bibfnamefont {T.}~\bibnamefont {Vieijra}},
  \bibinfo {author} {\bibfnamefont {J.}~\bibnamefont {Nys}}, \ and\ \bibinfo
  {author} {\bibfnamefont {J.}~\bibnamefont {Ryckebusch}},\ }\href {\doibase
  10.1103/PhysRevE.99.023304} {\bibfield  {journal} {\bibinfo  {journal} {Phys.
  Rev. E}\ }\textbf {\bibinfo {volume} {99}},\ \bibinfo {pages} {023304}
  (\bibinfo {year} {2019})}\BibitemShut {NoStop}%
\bibitem [{\citenamefont {Canabarro}\ \emph
  {et~al.}(2019{\natexlab{b}})\citenamefont {Canabarro}, \citenamefont
  {Fanchini}, \citenamefont {Malvezzi}, \citenamefont {Pereira},\ and\
  \citenamefont {Chaves}}]{PhysRevB.100.045129}%
  \BibitemOpen
  \bibfield  {author} {\bibinfo {author} {\bibfnamefont {A.}~\bibnamefont
  {Canabarro}}, \bibinfo {author} {\bibfnamefont {F.~F.}\ \bibnamefont
  {Fanchini}}, \bibinfo {author} {\bibfnamefont {A.~L.}\ \bibnamefont
  {Malvezzi}}, \bibinfo {author} {\bibfnamefont {R.}~\bibnamefont {Pereira}}, \
  and\ \bibinfo {author} {\bibfnamefont {R.}~\bibnamefont {Chaves}},\ }\href
  {\doibase 10.1103/PhysRevB.100.045129} {\bibfield  {journal} {\bibinfo
  {journal} {Phys. Rev. B}\ }\textbf {\bibinfo {volume} {100}},\ \bibinfo
  {pages} {045129} (\bibinfo {year} {2019}{\natexlab{b}})}\BibitemShut
  {NoStop}%
\bibitem [{\citenamefont {Wetzel}\ and\ \citenamefont
  {Scherzer}(2017)}]{WetzelSU2}%
  \BibitemOpen
  \bibfield  {author} {\bibinfo {author} {\bibfnamefont {S.~J.}\ \bibnamefont
  {Wetzel}}\ and\ \bibinfo {author} {\bibfnamefont {M.}~\bibnamefont
  {Scherzer}},\ }\href {\doibase 10.1103/PhysRevB.96.184410} {\bibfield
  {journal} {\bibinfo  {journal} {Phys. Rev. B}\ }\textbf {\bibinfo {volume}
  {96}},\ \bibinfo {pages} {184410} (\bibinfo {year} {2017})}\BibitemShut
  {NoStop}%
\bibitem [{\citenamefont {Wang}\ and\ \citenamefont
  {Zhai}(2017)}]{WangZhaiPCA1}%
  \BibitemOpen
  \bibfield  {author} {\bibinfo {author} {\bibfnamefont {C.}~\bibnamefont
  {Wang}}\ and\ \bibinfo {author} {\bibfnamefont {H.}~\bibnamefont {Zhai}},\
  }\href {\doibase 10.1103/PhysRevB.96.144432} {\bibfield  {journal} {\bibinfo
  {journal} {Phys. Rev. B}\ }\textbf {\bibinfo {volume} {96}},\ \bibinfo
  {pages} {144432} (\bibinfo {year} {2017})}\BibitemShut {NoStop}%
\bibitem [{\citenamefont {Wang}\ and\ \citenamefont
  {Zhai}(2018)}]{WangZhaiPCA2}%
  \BibitemOpen
  \bibfield  {author} {\bibinfo {author} {\bibfnamefont {C.}~\bibnamefont
  {Wang}}\ and\ \bibinfo {author} {\bibfnamefont {H.}~\bibnamefont {Zhai}},\
  }\href {\doibase 10.1007/s11467-018-0798-7} {\bibfield  {journal} {\bibinfo
  {journal} {Front. Phys.}\ }\textbf {\bibinfo {volume} {13}},\ \bibinfo
  {pages} {130507} (\bibinfo {year} {2018})}\BibitemShut {NoStop}%
\bibitem [{\citenamefont {Hu}\ \emph {et~al.}(2017)\citenamefont {Hu},
  \citenamefont {Singh},\ and\ \citenamefont {Scalettar}}]{SinghUMLXY}%
  \BibitemOpen
  \bibfield  {author} {\bibinfo {author} {\bibfnamefont {W.}~\bibnamefont
  {Hu}}, \bibinfo {author} {\bibfnamefont {R.~R.~P.}\ \bibnamefont {Singh}}, \
  and\ \bibinfo {author} {\bibfnamefont {R.~T.}\ \bibnamefont {Scalettar}},\
  }\href {\doibase 10.1103/PhysRevE.95.062122} {\bibfield  {journal} {\bibinfo
  {journal} {Phys. Rev. E}\ }\textbf {\bibinfo {volume} {95}},\ \bibinfo
  {pages} {062122} (\bibinfo {year} {2017})}\BibitemShut {NoStop}%
\bibitem [{\citenamefont {Wetzel}(2017)}]{WetzelAutoencoders}%
  \BibitemOpen
  \bibfield  {author} {\bibinfo {author} {\bibfnamefont {S.~J.}\ \bibnamefont
  {Wetzel}},\ }\href {\doibase 10.1103/PhysRevE.96.022140} {\bibfield
  {journal} {\bibinfo  {journal} {Phys. Rev. E}\ }\textbf {\bibinfo {volume}
  {96}},\ \bibinfo {pages} {022140} (\bibinfo {year} {2017})}\BibitemShut
  {NoStop}%
\bibitem [{\citenamefont {Zhang}\ \emph {et~al.}(2019)\citenamefont {Zhang},
  \citenamefont {Liu},\ and\ \citenamefont {Wei}}]{Zhang_percolation}%
  \BibitemOpen
  \bibfield  {author} {\bibinfo {author} {\bibfnamefont {W.}~\bibnamefont
  {Zhang}}, \bibinfo {author} {\bibfnamefont {J.}~\bibnamefont {Liu}}, \ and\
  \bibinfo {author} {\bibfnamefont {T.-C.}\ \bibnamefont {Wei}},\ }\href
  {\doibase 10.1103/PhysRevE.99.032142} {\bibfield  {journal} {\bibinfo
  {journal} {Phys. Rev. E}\ }\textbf {\bibinfo {volume} {99}},\ \bibinfo
  {pages} {032142} (\bibinfo {year} {2019})}\BibitemShut {NoStop}%
\bibitem [{\citenamefont {{Iwasaki}}\ \emph {et~al.}(2019)\citenamefont
  {{Iwasaki}}, \citenamefont {{Sawada}}, \citenamefont {{Stanev}},
  \citenamefont {{Ishida}}, \citenamefont {{Kirihara}}, \citenamefont
  {{Omori}}, \citenamefont {{Someya}}, \citenamefont {{Takeuchi}},
  \citenamefont {{Saitoh}},\ and\ \citenamefont {{Yorozu}}}]{2019npjCM5}%
  \BibitemOpen
  \bibfield  {author} {\bibinfo {author} {\bibfnamefont {Y.}~\bibnamefont
  {{Iwasaki}}}, \bibinfo {author} {\bibfnamefont {R.}~\bibnamefont {{Sawada}}},
  \bibinfo {author} {\bibfnamefont {V.}~\bibnamefont {{Stanev}}}, \bibinfo
  {author} {\bibfnamefont {M.}~\bibnamefont {{Ishida}}}, \bibinfo {author}
  {\bibfnamefont {A.}~\bibnamefont {{Kirihara}}}, \bibinfo {author}
  {\bibfnamefont {Y.}~\bibnamefont {{Omori}}}, \bibinfo {author} {\bibfnamefont
  {H.}~\bibnamefont {{Someya}}}, \bibinfo {author} {\bibfnamefont
  {I.}~\bibnamefont {{Takeuchi}}}, \bibinfo {author} {\bibfnamefont
  {E.}~\bibnamefont {{Saitoh}}}, \ and\ \bibinfo {author} {\bibfnamefont
  {S.}~\bibnamefont {{Yorozu}}},\ }\href {\doibase 10.1038/s41524-019-0241-9}
  {\bibfield  {journal} {\bibinfo  {journal} {npj Comput. Mater.}\ }\textbf
  {\bibinfo {volume} {5}},\ \bibinfo {eid} {103} (\bibinfo {year}
  {2019})}\BibitemShut {NoStop}%
\bibitem [{\citenamefont {Ponte}\ and\ \citenamefont
  {Melko}(2017{\natexlab{b}})}]{PhysRevB.96.205146}%
  \BibitemOpen
  \bibfield  {author} {\bibinfo {author} {\bibfnamefont {P.}~\bibnamefont
  {Ponte}}\ and\ \bibinfo {author} {\bibfnamefont {R.~G.}\ \bibnamefont
  {Melko}},\ }\href {\doibase 10.1103/PhysRevB.96.205146} {\bibfield  {journal}
  {\bibinfo  {journal} {Phys. Rev. B}\ }\textbf {\bibinfo {volume} {96}},\
  \bibinfo {pages} {205146} (\bibinfo {year} {2017}{\natexlab{b}})}\BibitemShut
  {NoStop}%
\bibitem [{\citenamefont {Wu}\ and\ \citenamefont
  {Zhai}()}]{wu2019generalized}%
  \BibitemOpen
  \bibfield  {author} {\bibinfo {author} {\bibfnamefont {Y.}~\bibnamefont
  {Wu}}\ and\ \bibinfo {author} {\bibfnamefont {H.}~\bibnamefont {Zhai}},\
  }\href {https://arxiv.org/abs/1904.05067v1} {\ }\Eprint
  {http://arxiv.org/abs/1904.05067} {arXiv:1904.05067} \BibitemShut {NoStop}%
\bibitem [{\citenamefont {Hou}\ \emph {et~al.}()\citenamefont {Hou},
  \citenamefont {Wong},\ and\ \citenamefont {Huang}}]{hou2019minimal}%
  \BibitemOpen
  \bibfield  {author} {\bibinfo {author} {\bibfnamefont {T.}~\bibnamefont
  {Hou}}, \bibinfo {author} {\bibfnamefont {K.~Y.~M.}\ \bibnamefont {Wong}}, \
  and\ \bibinfo {author} {\bibfnamefont {H.}~\bibnamefont {Huang}},\ }\href
  {https://arxiv.org/abs/1904.05067v1} {\ }\Eprint
  {http://arxiv.org/abs/1904.13052} {arXiv:1904.13052} \BibitemShut {NoStop}%
\bibitem [{\citenamefont {{Jadrich}}\ \emph {et~al.}(2018)\citenamefont
  {{Jadrich}}, \citenamefont {{Lindquist}},\ and\ \citenamefont
  {{Truskett}}}]{2018JChPh.149s4109J}%
  \BibitemOpen
  \bibfield  {author} {\bibinfo {author} {\bibfnamefont {R.~B.}\ \bibnamefont
  {{Jadrich}}}, \bibinfo {author} {\bibfnamefont {B.~A.}\ \bibnamefont
  {{Lindquist}}}, \ and\ \bibinfo {author} {\bibfnamefont {T.~M.}\ \bibnamefont
  {{Truskett}}},\ }\href {\doibase 10.1063/1.5049849} {\bibfield  {journal}
  {\bibinfo  {journal} {J. Chem. Phys.}\ }\textbf {\bibinfo {volume} {149}},\
  \bibinfo {eid} {194109} (\bibinfo {year} {2018})}\BibitemShut {NoStop}%
\bibitem [{\citenamefont {{Shah}}(2019)}]{2019JPhCo3g5006S}%
  \BibitemOpen
  \bibfield  {author} {\bibinfo {author} {\bibfnamefont {S.~N.}\ \bibnamefont
  {{Shah}}},\ }\href {\doibase 10.1088/2399-6528/ab3029} {\bibfield  {journal}
  {\bibinfo  {journal} {J. Phys. Commun.}\ }\textbf {\bibinfo {volume} {3}},\
  \bibinfo {pages} {075006} (\bibinfo {year} {2019})}\BibitemShut {NoStop}%
\bibitem [{\citenamefont {Mattis}(1976)}]{Mattis}%
  \BibitemOpen
  \bibfield  {author} {\bibinfo {author} {\bibfnamefont {D.}~\bibnamefont
  {Mattis}},\ }\href {\doibase https://doi.org/10.1016/0375-9601(76)90396-0}
  {\bibfield  {journal} {\bibinfo  {journal} {Physics Letters A}\ }\textbf
  {\bibinfo {volume} {56}},\ \bibinfo {pages} {421 } (\bibinfo {year}
  {1976})}\BibitemShut {NoStop}%
\bibitem [{\citenamefont {Fischer}\ and\ \citenamefont
  {Hertz}(1991)}]{FischerSpinGlass}%
  \BibitemOpen
  \bibfield  {author} {\bibinfo {author} {\bibfnamefont {K.~H.}\ \bibnamefont
  {Fischer}}\ and\ \bibinfo {author} {\bibfnamefont {J.~A.}\ \bibnamefont
  {Hertz}},\ }\href@noop {} {\emph {\bibinfo {title} {{Spin Glasses}}}}\
  (\bibinfo  {publisher} {Cambridge University Press},\ \bibinfo {year}
  {1991})\BibitemShut {NoStop}%
\bibitem [{\citenamefont {Pearson}(1901)}]{KarlPeason}%
  \BibitemOpen
  \bibfield  {author} {\bibinfo {author} {\bibfnamefont {K.}~\bibnamefont
  {Pearson}},\ }\href {\doibase 10.1080/14786440109462720} {\bibfield
  {journal} {\bibinfo  {journal} {Philos. Mag.}\ }\textbf {\bibinfo {volume}
  {2}},\ \bibinfo {pages} {559} (\bibinfo {year} {1901})}\BibitemShut {NoStop}%
\bibitem [{\citenamefont {Gingras}(1991{\natexlab{a}})}]{Michel1}%
  \BibitemOpen
  \bibfield  {author} {\bibinfo {author} {\bibfnamefont {M.~J.~P.}\
  \bibnamefont {Gingras}},\ }\href {\doibase 10.1103/PhysRevB.44.7139}
  {\bibfield  {journal} {\bibinfo  {journal} {Phys. Rev. B}\ }\textbf {\bibinfo
  {volume} {44}},\ \bibinfo {pages} {7139} (\bibinfo {year}
  {1991}{\natexlab{a}})}\BibitemShut {NoStop}%
\bibitem [{\citenamefont {Gingras}(1991{\natexlab{b}})}]{Michel2}%
  \BibitemOpen
  \bibfield  {author} {\bibinfo {author} {\bibfnamefont {M.~J.~P.}\
  \bibnamefont {Gingras}},\ }\href {\doibase 10.1103/PhysRevB.43.13747}
  {\bibfield  {journal} {\bibinfo  {journal} {Phys. Rev. B}\ }\textbf {\bibinfo
  {volume} {43}},\ \bibinfo {pages} {13747} (\bibinfo {year}
  {1991}{\natexlab{b}})}\BibitemShut {NoStop}%
\bibitem [{\citenamefont {Gingras}(1992)}]{Michel3}%
  \BibitemOpen
  \bibfield  {author} {\bibinfo {author} {\bibfnamefont {M.~J.~P.}\
  \bibnamefont {Gingras}},\ }\href {\doibase 10.1103/PhysRevB.45.7547}
  {\bibfield  {journal} {\bibinfo  {journal} {Phys. Rev. B}\ }\textbf {\bibinfo
  {volume} {45}},\ \bibinfo {pages} {7547} (\bibinfo {year}
  {1992})}\BibitemShut {NoStop}%
\bibitem [{Note1()}]{Note1}%
  \BibitemOpen
  \bibinfo {note} {It is easier to determine PCA's success with determining the
  gauge variables if they are taken from a discrete distribution, which can be
  clearly identified in a histogram such as Fig.~\ref {fig:extracted}, as
  opposed to a continuous distribution.}\BibitemShut {Stop}%
\bibitem [{\citenamefont {Nishimori}(2001)}]{NishimoriSpinGlass}%
  \BibitemOpen
  \bibfield  {author} {\bibinfo {author} {\bibfnamefont {H.}~\bibnamefont
  {Nishimori}},\ }\href@noop {} {\emph {\bibinfo {title} {{Statistical Physics
  of Spin Glasses and Information Processing: An Introduction}}}}\ (\bibinfo
  {publisher} {Oxford University Press},\ \bibinfo {year} {2001})\BibitemShut
  {NoStop}%
\end{thebibliography}%

\end{document}